\documentclass[10pt,a4paper]{iopartMod} 

\expandafter\let\csname equation*\endcsname\relax 
 \expandafter\let\csname endequation*\endcsname\relax 

\usepackage{amsmath,amssymb,float}
\usepackage{bm,mathbbol}
\usepackage{graphicx,xcolor}
\usepackage[colorlinks=true,citecolor=blue]{hyperref}
\usepackage{color,graphicx,geometry}

\newcommand{ \rmL }{ L }

\newcommand{ \calA }{ \mathcal{A} }
\newcommand{ \calB }{ \mathcal{B} }

\renewcommand{\Re}{\operatorname{Re}}
\renewcommand{\Im}{\operatorname{Im}}

\begin{document}

\title[]{Gravitational lensing beyond geometric optics: II. Metric independence}
\author{Abraham I. Harte}
\address{Centre for Astrophysics and Relativity, School of Mathematical Sciences
	\\
	Dublin City University, Glasnevin, Dublin 9, Ireland}

\ead{abraham.harte@dcu.ie}

\begin{abstract}
Typical applications of gravitational lensing use the properties of electromagnetic or gravitational waves to infer the geometry through which those waves propagate. Nevertheless, the optical fields themselves---as opposed to their interactions with material bodies---encode very little of that geometry: It is shown here that any given configuration is compatible with a very large variety of spacetime metrics. For scalar fields in geometric optics, or observables which are not sensitive to the detailed polarization content of electromagnetic or gravitational waves, seven of the ten metric components are essentially irrelevant. With polarization, five components are irrelevant. In the former case, this result together with diffeomorphism invariance allows essentially any geometric-optics configuration associated with a particular spacetime to be embedded into any other spacetime, at least in finite regions. Going beyond the geometric-optics approximation breaks some of this degeneracy, although much remains even then. Overall, high-frequency wave propagation is shown to be insensitive to compositions of certain conformal, Kerr-Schild, and related transformations of the background metric. One application is that new solutions for scalar, electromagnetic, and gravitational waves may be generated from old ones. In one example described here, the high-frequency scattering of a plane wave by a point mass is computed by transforming a plane wave in flat spacetime.
\end{abstract}

\section{Introduction}

Historically, the confirmation that starlight is bent at the predicted angle by the gravitational field of the Sun was one of the most important milestones in the acceptance of general relativity as a physical theory. Today, the situation is in some sense reversed: General relativity is used less to predict the lensing associated with known masses, and more to infer the properties of unseen matter in terms of its lensing effects \cite{SchneiderEhlers, BartelmannRev, darkMatter}. Although these inferences are rarely described as such, they are solutions to inverse problems. Given that certain fields are expected to solve a particular class of metric-dependent equations, and given (partial) knowledge of one or more particular solutions, they ask what might be learned about the underlying metric, and thence about the matter which sources that metric. In practice, it is typical to solve such problems by producing a parametrized family of models, computing their associated observables, and then fitting the free parameters to the data.

This approach requires some care. Inverse problems are often ill-posed \cite{InverseProbs}, so solutions may fail to be unique or even to exist. Worse,  solutions can depend very sensitively, or even discontinuously, on the observational inputs. Although difficulties such as these afflict a wide variety of inverse problems, one might initially hope that they do not do so in the context considered here. The laws of geometric optics typically employed in gravitational lensing make essential use of the geometry, suggesting that much of that geometry can be inferred from optical measurements. Indeed, knowledge of all null geodesics does fix the metric up to a conformal factor, and supplementing this with knowledge of intensity variations fixes even that. Nevertheless, realistic observations involve only \textit{some} null geodesics, and even those are characterized (at best) at their observation and emission events, but not in between. This lack of information greatly enlarges the number of geometries which might be compatible with any given set of observations.

This paper does not attempt to discuss the interpretation of realistic astronomical data, but instead works towards answering a related but more fundamental question: How does wave propagation depend on the underlying geometry? Or rather, how does it not? We ask which quantities can be preserved when the metric changes. This is similar to asking for the symmetries of the optical equations, but instead of finding operations which preserve \textit{all} solutions, we look for those which preserve a \textit{particular} solution. This distinction allows for a much broader class of possibilities---possibilities which might be described as ``solution-dependent symmetries.'' More precisely, we characterize those ways in which metrics can be transformed without affecting a particular optical field---in geometric optics or beyond, and for scalar, electromagnetic, and gravitational waves. This helps to determine how much of the geometry can really be inferred from an optical field. It may also be viewed as a way to characterize the kernel of an inverse problem which asks for the metric given a set of optical observables.

Similar questions have been addressed in an exact setting at least for electromagnetic fields. It is well-known that if $F_{ab}$ is an exact solution to the source-free Maxwell equations on a background with metric $g_{ab}$, it is also an exact solution on all conformally-related backgrounds $\Omega^2 g_{ab}$ \cite{BatemanConf, Wald, Kastrup2008}. Indeed, there is a sense in which knowledge of \textit{all} Maxwell solutions determines the metric up to the single free function $\Omega$.  However, if only a \textit{single} $F_{ab}$ is known, much less can be determined; five of the ten metric components are essentially irrelevant \cite{HarteEMKS}. This additional (non-conformal) freedom may be associated with Kerr-Schild and related transformations generated by the principal null directions of the electromagnetic field. 

Similar results are not known for scalar or gravitational fields. While it may be possible to find them using similar techniques, one rarely has access to exact solutions even in the electromagnetic case. Instead, it is more common---both theoretically and observationally---to rely on observables which are associated with a high-frequency, or ``geometric optics,'' limit. The approach here focuses on exactly this limit. We consider cases in which solutions to the equations of geometric optics and its first corrections remain solutions under a variety of metric transformations. This has the advantage that it is straightforward to consider fields which are not necessarily electromagnetic. However, even in the electromagnetic case, restricting to the high-frequency limit allows for stronger results than are obtained in an exact setting; approximate solutions depend even less on the geometry than exact ones.

This paper may be viewed as a continuation of two others. The first \cite{HarteEMKS} derived the aforementioned metric invariance of exact electromagnetic fields. The second \cite{HarteOptics1} is referred to below as Paper I, and set out the foundations of high-frequency wave propagation for Klein-Gordon, electromagnetic, and gravitational waves in general relativity. The layout here is as follows: Sect. \ref{Sect:Review} reviews the main equations of geometric optics and its corrections, as derived in Paper I and elsewhere. Sect. \ref{Sect:Rays} discusses metric invariance for the null geodesics which play a central role in the propagation of high-frequency fields. The primary metric transformations used in the remainder of the paper are obtained here. Sect. \ref{Sect:Geo} discusses how those metric transformations preserve not only the rays of geometric optics, but also the amplitudes. Consequent transformations of various observables are obtained as well. In Sect. \ref{Sect:Inheritance}, it is explained how invariance results in geometric optics imply invariance results at least for certain quantities beyond geometric optics. Sect. \ref{Sect:BeyondGO} fills in further details at one order beyond geometric optics, considering how metric transformations affect the subleading amplitudes---particularly in cases where they induce only phase shifts. Sects. \ref{Sect:Examples} provides some examples which illustrate how these results can be used to generate new solutions from old ones using purely algebraic operations. It is shown, e.g., that spherical waves are universal in spherically-symmetric metrics and plane-fronted waves are universal in plane-fronted geometries. Sect. \ref{Sect:Scattering} discusses a more involved application, where a plane wave scattered by a point mass is obtained from a plane wave in flat spacetime. There are two appendices. \ref{app:Metric} discusses geometrical properties of the metric transformations which are found to preserve the optical fields. \ref{app:Notation} explains notation and conventions, and includes tables which index many of the symbols used in the paper.

\section{A summary of high-frequency approximations}
\label{Sect:Review}

As discussed in more detail in paper I and elsewhere \cite{BornWolf, Keller, EhlersGeoOptics, Anile, Isaacson1, DolanGeoOptics1, ThorneBlandford}, geometric optics and its corrections may be derived by considering 1-parameter families of fields with the form\footnote{This expansion is intended only to be asymptotic. For fixed finite $\omega$, the right-hand side does not generically converge to the left-hand side (and may not converge at all). As usual for asymptotic series, a finite truncation of the series on the right-hand side should be viewed as an approximation for the left-hand side as $\omega \to \infty$. Nevertheless, equality symbols are used here and below for simplicity.}
\begin{equation}
	\psi_B (x;\omega) = e^{i \omega \varphi(x) } \sum_{n=0}^\infty \omega^{-n} \calA^n_B(x),
	\label{WKB}
\end{equation}
where $\varphi$ is called the phase (or eikonal) function, the $\calA^n_B$ are referred to as amplitudes, and $\omega$ is a large parameter which controls the frequency of the solution. The multi-index $B$ stands for a collection of indices $b_1 \cdots b_s$ appropriate to the particular field under consideration. For scalar fields, $s=0$; for electromagnetic vector potentials, $s=1$; for metric perturbations, $s=2$. The fields $\psi_B$ in these three cases were respectively denoted by $\psi$, $a_a$, and $h_{ab}$ in Paper I. Although they are complex, only their real components
\begin{equation}
	\Psi_B \equiv \Re \psi_B
	\label{RePsi}
\end{equation}
are to be considered as physical.

The eikonal and the amplitudes may be determined by inserting the ansatz \eqref{WKB} for the potentials into the appropriate gauge-fixed field equation. This is assumed here to be of the form
\begin{equation}
	\mathcal{D} \psi_B \equiv \big( \delta^A_B g^{cd} \nabla_c \nabla_d - \mathcal{M}^{A}{}_{B} \big) \psi_A = 0,
	\label{fieldEqn}
\end{equation}
where $\mathcal{M}^{A}{}_{B}$ is an ordinary tensor field. For the Klein-Gordon equation with curvature coupling $\xi$ and field mass $\mu$, Maxwell's equations in Lorenz gauge, and the linearized Einstein equation in Lorenz gauge,
\begin{gather}
\mathcal{M} = \xi R + \mu^2, \qquad \mathcal{M}^{a}{}_{b} = g^{ac} R_{bc}, \qquad \mathcal{M}^{ab}{}_{cd} = 2 g^{ae} R_{ecd}{}^{b},
\label{Mdef}
\end{gather}
where $R_{abc}{}^{d}$ is the Riemann tensor, $R_{ab} = R_{acb}{}^{c}$ the Ricci tensor, $R = g^{ab} R_{ab}$ the Ricci scalar, and $\nabla_a$ the covariant derivative associated with the metric $g_{ab}$. 

Regardless of the particular form of $\mathcal{M}^{A}{}_{B}$, combining \eqref{WKB} with \eqref{fieldEqn} shows that $\varphi$ must satisfy the eikonal equation
\begin{equation}
	g^{ab} k_a k_b = 0, \qquad k_a \equiv - \nabla_a \varphi,
	\label{eikonal}
\end{equation}
which implies that hypersurfaces of constant $\varphi$ must be null; these are the hypersurfaces of constant phase in the $\omega \to \infty$ limit. Equivalently, \eqref{eikonal} implies that the wave(co)vector $k_a$ is an exact null covector. Raising an index, $g^{ab} k_b$ is a null, geodesic, and twist-free vector field; its integral curves are the rays of geometric optics. 

The field equation \eqref{fieldEqn} constrains not only the eikonal, but also the amplitudes $\calA_B^n$. In terms of the transport operator
\begin{equation}
	\rmL \equiv g^{ab} ( 2 k_a \nabla_b + \nabla_a k_b),
	\label{LDef}
\end{equation}
which may be viewed as an ordinary differential operator along the rays, the amplitudes satisfy the transport equations
\begin{equation}
	\rmL \calA^n_B = - i \mathcal{D} \calA^{n-1}_B.
	\label{xPort}
\end{equation}
The $n=0$ case of this equation is sometimes written separately as $\rmL \calA^0_B = 0$, although we avoid this here by defining $\calA^{-1}_B \equiv 0$; Eqn. \eqref{xPort} then holds for all $n \geq 0$. Regardless, the transport equations are effectively differential constraints on the amplitudes. The $n=0$ case corresponds to geometric optics, and the lack of any source term implies that each ray evolves independently of its neighbors. For $n > 0$, the $- i \mathcal{D} \calA^{n-1}_B$ source provides a coupling between neighboring rays. Roughly speaking, it may be interpreted as accounting for interference between those rays.

If $s=0$, the eikonal and transport equations are the only consequences of combining the high-frequency ansatz with the Klein-Gordon equation. If $s > 0$ however, these equations must be supplemented with additional constraints which arise as a consequence of the gauge conditions used to reduce Maxwell's equations or the linearized Einstein equation to the form \eqref{fieldEqn}. In the $s=1$ case of an electromagnetic vector potential, the Lorenz gauge requires that
\begin{equation}
	g^{ab} (\mathcal{A}_{a}^n k_b + i \nabla_a \mathcal{A}^{n-1}_{b}) = 0.
	\label{gaugeEM}
\end{equation}
If $s=2$, the Lorenz gauge applied to metric perturbations instead requires that
\begin{equation}
	(\delta^b_a g^{cd} - \tfrac{1}{2} \delta^d_a g^{bc}) (\mathcal{A}_{bc}^n k_d + i \nabla_d \mathcal{A}_{bc}^{n-1}) = 0.
	\label{gaugeGrav}
\end{equation}
These constraints are algebraic. If they are satisfied on an initial hypersurface which is transverse to the rays, the transport equations guarantee that they remain satisfied away from that hypersurface.

Geometric optics is associated with knowledge only of $\varphi$ and the zeroth-order amplitude $\calA^0_B$. Supplementing these quantities with a metric $g_{ab}$ for which they are known to satisfy the eikonal, transport, and gauge equations, it is convenient to refer to the set $O_0 \equiv \{ g_{ab} ; \varphi, \calA^0_B\}$ as the ``0th-order optical fields.'' Generalizing this concept beyond geometric optics, let
\begin{equation}
	O_n \equiv \{ g_{ab}; \varphi, \calA^0_B, \ldots, \calA^n_B\}
	\label{ODef}
\end{equation}
denote the $n$th-order optical fields. Many of our main results below take the form $O_n \mapsto \hat{O}_n$, where knowledge of the optical fields associated with some seed metric $g_{ab}$ may be used to immediately write down optical fields associated with a transformed metric $\hat{g}_{ab}$.

As a brief note on applicability, the eikonal equation and the $n=0$ transport and gauge conditions which govern $O_0$ may be shown to imply the three ``laws of geometric optics'' which are the basis for much of the theory of gravitational lensing:
\begin{enumerate}
	\item Fields propagate along null geodesic rays.
	\item Intensities satisfy the area-intensity conservation law $\nabla_a (|\calA_0|^2 g^{ab} k_b) = 0$.
	\item Polarization tensors are parallel transported along the rays.
\end{enumerate}
If these laws are applied when $\omega$ is large but finite, there is a lengthscale $\ell$ such that errors in a predicted observable have relative magnitudes proportional to positive powers of the dimensionless number $(\omega \ell)^{-1}$. It is sometimes difficult to estimate $\ell$ without a detailed calculation, as it can have contributions from lengthscales associated with the spacetime curvature, wavefront curvature, amplitude gradient, shear scale, or some (possibly nonlocal) composite thereof; its precise nature depends on the details of the system under consideration and the particular observable which is considered. Nevertheless, there are some cases---such as for plane waves in flat spacetime---where there is no intrinsic scale and the leading-order high-frequency approximation is in fact exact\footnote{Interestingly, certain gravitational plane wave solutions to the geometric-optics equations are exact solutions not only to the linearized Einstein equation, but also to the exact \textit{nonlinear} Einstein equation \cite{HarteOptics1}. This occurs for solutions expressed in the Kerr-Schild form $\calA^0_{ab} \propto k_a k_b$, and is a consequence of the general result \cite{Guerses, Xanthopoulos1978, ExactSolns} that Einstein's equation is linear for Kerr-Schild perturbations when $k_a$ is geodesic.}. It may also be expected that as one moves away from a compact source in an asymptotically-flat spacetime, $\ell$ eventually increases without bound. \textit{Any} finite $\omega$ is thus ``large'' at sufficiently large distances, implying that geometric optics is a good approximation far from compact sources. If $(\omega \ell)^{-1}$ is small but not too small, corrections to geometric optics associated with, e.g., $O_1$ may nevertheless be measurable, and may provide information which is complementary to that supplied by geometric optics alone. Subleading corrections to various observables are discussed in Paper I.

\section{Metric invariance of the optical rays}
\label{Sect:Rays}

The most basic quantity which appears in geometric optics is the eikonal $\varphi$. It is thus natural to begin a discussion of metric invariance in geometric optics by discussing metric invariance for the eikonal equation \eqref{eikonal}: If a particular $\varphi$ is known to be a solution to that equation with the metric $g_{ab}$, for which metrics $\hat{g}_{ab}$ does $\varphi$ satisfy the transformed eikonal equation $\hat{g}^{ab} \nabla_a \varphi \nabla_b \varphi = 0$? This constitutes only a single scalar constraint on the inverse of the transformed metric, so only one of the ten scalar functions which comprise it is constrained. The general class of transformations $g_{ab} \mapsto \hat{g}_{ab}$ which preserve a given eikonal involve nine free functions, and in this sense, knowledge of an eikonal alone is locally compatible with almost any metric whatsoever. 

While it is sometimes useful to consider metric transformations which preserve eikonals alone, these do not necessarily preserve any other interesting aspects of geometric optics. Much more can be said by considering the somewhat smaller class of metric transformations which preserve both an eikonal \textit{and} its associated rays. Fixing $k_a = - \nabla_a \varphi$, this is equivalent to demanding that $g^{ab} k_b$ be proportional to $\hat{g}^{ab} k_b$, or that $k_c$ be an eigenvector of $\hat{g}_{ab} g^{bc}$. Hence,
\begin{equation}
	k_{[a} (\hat{g}_{b]c} g^{cd} k_d) = 0,
	\label{kEig}
\end{equation}
which may be interpreted as three scalar constraints. The transformations $g_{ab} \mapsto \hat{g}_{ab}$ which preserve a given eikonal and its rays may thus be expected to involve seven free functions.

These transformation can be written in a variety of ways. Here we do so by first fixing a complex null tetrad
\begin{equation}
	(k_a, n_a, m_a, \bar{m}_a)
	\label{tetrad}
\end{equation}
in which $n_a$ is real, $\bar{m}_a$ is the complex conjugate of $m_a$, and the only non-vanishing inner products with respect to $g_{ab}$ are
\begin{equation}
	g^{ab} m_a \bar{m}_b = - g^{ab} k_a n_b = 1.
	\label{tetradNorm}
\end{equation}	
Now, any rank-2 symmetric tensor may be expanded as a linear combination of symmetrized products of the tetrad components. Performing such an expansion for $\hat{g}_{ab}$ while imposing \eqref{kEig} and requiring that the result be real implies that 
\begin{equation}
	\hat{g}_{ab} =  [ r_1 k_{(a} + r_2 n_{(a} + (\bar{c}_1  m_{(a} + c_1 \bar{m}_{(a} )] k_{b) } + \bar{c}_2 m_a m_b + c_2 \bar{m}_a \bar{m}_b + r_3 m_{(a} \bar{m}_{b)},
	\label{Hexpand}
\end{equation}
where $r_1, r_2, r_3$ are real and $c_1, c_2$ may be complex. Any collection of such scalars is allowed as long as the resulting metric is non-singular and has the same signature as $g_{ab}$. Although this expansion for $\hat{g}_{ab}$ is simple to write down, the various scalars which appear in it do not have particularly transparent interpretations. For this reason and others, it is difficult to organize calculations in which geometric structures associated with $\hat{g}_{ab}$ are expressed in terms of ``background'' structures associated with $g_{ab}$.

These problems are considerably alleviated by replacing the above scalars with a real conformal factor $\Omega$, a real covector $w_a$, and a complex scalar $Y$, such that
\begin{equation}
	\hat{g}_{ab} = \Omega^2 \left( g_{ab} + k_{(a} w_{b)} + \frac{ \bar{Y} m_a m_b + |Y|^2 m_{(a} \bar{m}_{b)} + Y \bar{m}_a \bar{m}_b }{ 1 - \frac{1}{4} |Y|^2 }  \right) .
	\label{gTilde}
\end{equation}
Together, $\Omega$, $w_a$, and $Y$ constitute seven real ``deformation functions.'' Using $g_{ab} = 2 (m_{(a} \bar{m}_{b)} - k_{(a} n_{b)})$, they may be shown to be related to the scalars in \eqref{Hexpand} via
\begin{equation}
\begin{gathered}
	\Omega^4 = \tfrac{1}{4} r_3^2 - |c_2|^2, \qquad Y = \frac{ 4 c_2 }{ r_3 + 2 \Omega^2 },
	\\
	w_a = \Omega^{-2} \left[ r_1 k_a + ( r_2 + 2 \Omega^2 ) n_a + \bar{c}_1 m_a + c_1 \bar{m}_a \right] . 
\end{gathered}
\end{equation}
Regardless, if Einstein's equation is not imposed, $\Omega$, $w_a$, and $Y$ are almost completely free; cf. \eqref{badParams}. Moreover, it is shown in \ref{app:Metric} that $\hat{g}_{ab}$ may be viewed as the composition of two complex Kerr-Schild transformations generated by $Y$ and $m_a$, a real extended Kerr-Schild transformation generated by $k_a$ and $w_a$, and a conformal transformation generated by $\Omega$. The simplicity of each of these individual transformations allows, e.g., the inverse metric $\hat{g}^{ab}$ and the volume element of $\hat{\epsilon}_{abcd}$ to be found exactly in terms of their background counterparts. Using the inverse \eqref{gInvHat}, one consequence is that 
\begin{equation}
	\hat{g}^{ab} k_b = \left( \frac{ \Omega^{-2}  }{ 1 + \tfrac{1}{2} k \cdot w } \right) g^{ab} k_b ,
	\label{kHat}
\end{equation}
which verifies that as claimed, the optical rays are preserved. As these rays are twist-free by construction, they are not only null, but also geodesic with respect to both $g_{ab}$ and $\hat{g}_{ab}$. Such characteristics appear to be the minimal ones in which geometric optics might be expected to be ``mostly preserved.''

Special cases of transformations $g_{ab} \mapsto \hat{g}_{ab}$ with the form \eqref{gTilde} have already been considered in a number of different contexts. Most obviously, those in which $w_a = Y = 0$ are conformal. If instead $\Omega=1$, $Y=0$, and $w_{[a} k_{b]} = 0$, they are Kerr-Schild transformations associated with $k_a$. If $\Omega = 1$ and $w_a = 0$ but $Y \neq 0$, they represent the real result of two complex Kerr-Schild transformations associated with null covectors transverse to $k_a$. In general, Kerr-Schild transformations are known to eliminate the nonlinearity in Einstein's equation, at least when the generating covector is geodesic \cite{Guerses, Xanthopoulos1978, ExactSolns}. Metrics associated with Kerr-Newman black holes, gravitational plane waves, (anti) de Sitter, and other important geometries can be written as Kerr-Schild transformations applied to flat backgrounds \cite{ExactSolns}. Furthermore, the Penrose limit \cite{PenroseLimit,Blau} implies that the geometry sufficiently near any null geodesic is a plane wave, suggesting that Kerr-Schild transformations applied to flat metrics are generically relevant in ultrarelativistic limits. If $Y=0$ and $w_a$ is null but not necessarily proportional to $k_a$, the transformations considered here fall into the class of extended Kerr-Schild transformations considered in \cite{Llosa2009, HarteXKS}; there is a sense in which these too ``reduce'' the nonlinearity of Einstein's equation. More general transformations in which $Y = 0$ are central to the ``optical geometry'' discussed in \cite{OptGeo}; these exactly preserve null (vacuum or force-free) Maxwell fields with principal null direction $k_a$ \cite{HarteEMKS}. Lastly, even if $Y \neq 0$, there is considerable overlap between the metric transformations considered here and those associated with the Ehlers group, which maps certain exact solutions of Einstein's equation to other exact solutions \cite{MarsEhlersGroup}. More specific examples involving physically-interesting pairs of metrics related by transformations $g_{ab} \mapsto \hat{g}_{ab}$ are discussed in Sect. \ref{Sect:Examples} below.

One interpretation of our general result that seven free functions may be used to  deform the metric without disturbing the optical rays is that only three of the ten components of $g_{ab}$ can be constrained by knowledge of any given twist-free null congruence. However, there is a sense in which even these remaining components are not essential; they may be ``gauged away.'' Given any twist-free null congruence associated with $g_{ab}$, and \textit{any} other well-behaved metric $\tilde{g}_{ab}$, there exist (at least in finite regions) diffeomorphisms $\phi$ such that the pullback $(\phi_* \tilde{g})_{ab}$ admits the same null congruence. Alternatively, these diffeomorphisms may be applied to push the rays---instead of the metric---forward so that they are compatible with $\tilde{g}_{ab}$ itself. In the first of these points of view, the eigenvector relation \eqref{kEig} is replaced by
\begin{equation}
	k_{[a} \left[ (\phi_* \tilde{g})_{b]c} g^{cd} k_d \right] = 0.
	\label{diffeo1}
\end{equation}
In the second, it is more suggestive to write
\begin{equation}
(\phi^{*} k)_{[a} \left[ \tilde{g}_{b]c} (\phi^* g^{-1}k )^{c} \right]  = 0. 
	\label{diffeo2}
\end{equation}
Both of these equations are however equivalent.

One corollary is that any system of optical rays associated with a given metric is also a valid system of rays associated with a flat metric: If $\tilde{g}_{ab}$ is flat and chosen arbitrarily, it is possible to find a diffeomorphism $\phi$ such that $\phi_* \tilde{g}_{ab}$ can be identified with $\hat{g}_{ab}$ and related to $g_{ab}$ via \eqref{gTilde}. The rays of $g_{ab}$ are thus rays also of the flat metric $\phi_* \tilde{g}_{ab}$. If inertial coordinates $x^\mu$ are chosen such that $\tilde{g}_{\mu\nu} \rmd x^\mu \rmd x^\nu = -(\rmd x^0)^2 + (\rmd x^1)^2 + (\rmd x^2)^2 + (\rmd x^3)^2$, the pushed-forward rays thus appear as straight lines\footnote{It is not uncommon to adopt coordinates in which the null geodesics emanating from a given timelike worldline appear to be straight. In the case of future-directed null geodesics, the retarded coordinates described in, e.g., \cite{PoissonRev} are one example. Closely related constructions, such as observational coordinates \cite{Ellis1985} and geodesic light cone coordinates \cite{Gasperini2011, Fleury2016}, are sometimes used in cosmology, where the preferred rays are associated with an observer's past-directed null geodesics.}, even though they are geodesics of the (potentially curved) metric with components $(\phi^* g)_{\mu\nu}$. These observations suggest that the rays alone---or at least a single collection of them---imply nothing about the local curvature.

The resulting ``universality'' of twist-free null congruences may be understood intuitively in terms of the freedom to identify points in one spacetime with points in another. This freedom can always be used to map twist-free null geodesics in one spacetime onto twist-free null geodesics in the other spacetime. What is perhaps more surprising is that it is shown in Sect. \ref{Sect:Geo} that the leading-order amplitudes also behave very simply under these mappings, a fact which is used explicitly in the scattering example considered in Sect. \ref{Sect:Scattering}.

\section{Amplitudes in geometric optics}
\label{Sect:Geo}

Suppose that a particular set of zeroth-order optical fields $O_0 = \{g_{ab}; \varphi, \calA_B^0 \}$ is known to solve the equations of geometric optics. Rewriting \eqref{eikonal} and \eqref{xPort}, it is thus assumed that
\begin{equation}
	g^{ab} \nabla_a \varphi \nabla_b \varphi = 0 , \qquad \rmL \calA^0_B = 0,
	\label{O0}
\end{equation}
where the transport operator $\rmL$ is defined by \eqref{LDef}. The gauge conditions \eqref{gaugeEM} or \eqref{gaugeGrav} are assumed to be satisfied as well if applicable. It then follows from the discussion in Sect. \ref{Sect:Rays} that if $\hat{g}_{ab}$ is obtained from $g_{ab}$ via a transformation with the form \eqref{gTilde}, $\varphi$ remains a valid eikonal in the transformed geometry. There thus exist valid collections of transformed fields with the form $\hat{O}_0 = \{ \hat{g}_{ab}; \varphi, \hat{\calA}_B^0 \}$, where $\hat{\rmL} \hat{\calA}_B^0 = 0$ and $\hat{\rmL} = \hat{g}^{ab} ( 2 k_a \hat{\nabla}_b + \hat{\nabla}_a k_b)$. We now complete this picture by deriving explicit transformation rules $\calA_B^0 \mapsto \hat{\calA}^0_B$.

\subsection{Geometric optics for scalar fields}

It is simplest to begin by considering transformations $\calA_0 \mapsto \hat{\calA}_0$ of the $s=0$ scalar amplitudes. Although these amplitudes directly govern the dynamics of high-frequency solutions to the Klein-Gordon equation, they are also relevant for electromagnetic and gravitational waves: Amplitudes in those cases may be factorized such that \cite{HarteOptics1}
\begin{equation}
	\calA^0_a = \calA_0 e_a , \qquad \calA^0_{ab} = \calA_0 e_{ab},
	\label{A0s}
\end{equation}
where the polarization tensors $e_a$ and $e_{ab} = e_{(ab)}$ are parallel-transported along the rays and satisfy the gauge conditions
\begin{equation}
	g^{ab} k_a e_b = 0, \qquad g^{ab} k_a (\delta^c_b \delta^d_f - \tfrac{1}{2} g_{ bf } g^{cd} ) e_{cd} = 0.
	\label{gauge0}
\end{equation}

Regardless of application, it follows from \eqref{O0} that the background amplitude $\calA_0$ must be a solution to $\rmL \calA_0 = 0$. An amplitude $\hat{\calA}_0$ which is associated with a transformed metric $\hat{g}_{ab}$ instead satisfies $\hat{\rmL} \hat{\calA}_0 =0$, so relating $\hat{\calA}_0$ to $\calA_0$ requires understanding how $\hat{\rmL}$ differs from $\rmL$. First note that for any $\hat{g}_{ab}$ with the form \eqref{gTilde}, it follows from \eqref{kHat} and \eqref{divHat} that 
\begin{align}
	\hat{\nabla}_a ( \hat{g}^{ab} k_b) = \frac{ \nabla \cdot k + 2 \mathcal{L}_k \ln \Omega  }{ \Omega^2 (1 + \tfrac{1}{2} k \cdot w) }  ,
\end{align}
where $\nabla \cdot k \equiv g^{ab} \nabla_a k_b$ and $k \cdot w \equiv g^{ab} k_a w_b$. The transformed transport operator applied to any scalar field $\calB$ thus satisfies
\begin{equation}
	\hat{\rmL} (\calB/\Omega) = \frac{ \rmL \calB }{ \Omega^3 (1 + \tfrac{1}{2} k 
	\cdot w) }.
	\label{Ltilde}
\end{equation}
It follows immediately that if $\calA_0$ is a valid geometric-optics amplitude associated with $g_{ab}$,
\begin{equation}
	\hat{\calA}_0 = \calA_0/\Omega	
	\label{A0scalarXform}
\end{equation}
is a valid amplitude associated with $\hat{g}_{ab}$.

Metric transformations generated by arbitrary $\Omega$, $w_a$, and $Y$ may thus be associated with optical fields which transform via\footnote{This transformation is not unique. In the absence of any initial conditions or similar constraints, the optical equations admit an infinite number of solutions. The result quoted here simply provides \textit{one} solution associated with the transformed metric, chosen for being simply related to a known solution in the background metric.} 
\begin{equation}
	O_0 = \{ g_{ab}; \varphi, \calA_0 \} \mapsto \hat{O}_0 = \{ \hat{g}_{ab}; \varphi, \Omega^{-1} \calA_0 \}.
	\label{mapScalar}
\end{equation}
Note that although there are seven deformation functions involved in the relation between $\hat{g}_{ab}$ and $g_{ab}$, only one of these---the conformal factor---can modify scalar fields in geometric optics; deformations associated with $w_a$ and $Y$ do not affect it. If one is interested in a metric $\tilde{g}_{ab}$ which cannot be generated using any combination of deformation functions, it follows from the discussion at the end of Sect. \ref{Sect:Rays} that there nevertheless exist diffeomorphisms $\phi$ such that $\phi_* \tilde{g}_{ab}$ \textit{can} be so generated (and thus identified with $\hat{g}_{ab}$).

Temporarily leaving aside any discussion of diffeomorphisms, it follows from \eqref{mapScalar} and \eqref{WKB} that to leading nontrivial order, gradients of scalar fields  are preserved up to conformal rescalings by any transformation $O_0 \mapsto \hat{O}_0$. More precisely,
\begin{align}
	\nabla_a \hat{\psi} &= \Omega^{-1} \nabla_a \psi + \mathcal{O}(\omega^{0} ) 
	\nonumber
	\\
	&= \omega \left[ - i\Omega^{-1} \calA_0 k_a e^{i \omega \varphi} + \mathcal{O}(\omega^{-1}) \right].
	\label{dpsi0}
\end{align}
Both $\psi$ and $\nabla_a \psi$ thus obey simple transformation rules. However, some observables which depend on them do not. For example, although the (complexified) forces $f_a = q \nabla_a \psi$ which act on charged test particles------rates of change of their momenta---are preserved up to scale at this order, the accelerations of those particles are not similarly preserved. To see this, first recall that for a test particle with mass $m$, charge $q$ and 4-velocity $u^a$,
\begin{equation}
	\frac{\mathrm{D} u^a}{\rmd \tau} = \frac{q}{m} ( g^{ab} + u^a u^b )  \nabla_b \Psi,
	\label{accScalar}
\end{equation}
where $\tau$ denotes a proper time along the particle's worldline and $\Psi = \Re \psi$. The projection operator $g^{ab} + u^a u^b$ is required here because there may be a component of the force vector $q g^{ab} \nabla_b \Psi$ which lies along $u^a$, thus changing a particle's mass. Regardless, in order to see how accelerations transform as $g_{ab} \mapsto \hat{g}_{ab}$, first suppose that $\hat{g}_{ab} u^a u^b < 0$ so the background 4-velocity is timelike with respect to both $g_{ab}$ and $\hat{g}_{ab}$. Then,
\begin{equation}
	\hat{u}^a \equiv \frac{u^a}{ \sqrt{ - \hat{g}_{bc} u^b u^c } } 
	\label{uXform}
\end{equation}
has unit norm with respect to $\hat{g}_{ab}$ and is a natural candidate for an ``equivalent'' 4-velocity in the hatted geometry. Transforming the right-hand side of \eqref{accScalar} using this and \eqref{kHat} and \eqref{dpsi0}, the hatted acceleration may be seen to reduce to
\begin{equation}
	\frac{ \hat{\mathrm{D}} \hat{u}^a }{ \rmd \hat{\tau} } = \frac{q}{m} \left[ \frac{g^{ab} }{ \Omega^3 ( 1 + \frac{1}{2} k \cdot w) } - \frac{ u^a u^b }{ \hat{g}_{cd} u^c u^d }  \right] \nabla_b \Psi + \mathcal{O}(\omega^{0}).
\end{equation}
Except in special cases, this is not proportional to $\mathrm{D} u^a/\rmd\tau$. Nevertheless, the difference between $\hat{\mathrm{D}} \hat{u}^a / \rmd \hat{\tau}$ and $[\Omega^3 ( 1 + \frac{1}{2} k \cdot w)]^{-1} \mathrm{D} u^a/\rmd\tau$ is proportional only to $u^a$. It arises because, although $u^a$ and $\mathrm{D} u^a/\rmd\tau$ are necessarily orthogonal with respect to $g_{ab}$, they might not be orthogonal with respect to $\hat{g}_{ab}$. In this example and more generally, observables which involve couplings to material bodies do not behave as simply under metric transformations as observables constructed from the optical fields alone.

An example of an observable which is constructed without reference to any material bodies is the average $\langle T_{ab} \rangle$ of a field's stress-energy tensor. This is defined and computed in Paper I in terms of a high-frequency expansion with the form
\begin{equation}
	\langle T_{ab}(x;\omega) \rangle = \frac{\omega^2 }{ 8\pi} \sum_{n=0}^\infty \omega^{-n} \mathcal{T}^n_{ab}(x).
	\label{Texpand}
\end{equation}
Each $\mathcal{T}^n_{ab}$ appearing here is separately conserved in the sense that $g^{ab} \nabla_a \mathcal{T}^n_{bc} = 0$. The $n=0$ component $\mathcal{T}^0_{ab} = |\calA_0|^2 k_a k_b$ depends only on the magnitude $|\calA_0|^2$, and it follows from \eqref{mapScalar} that
\begin{equation}
	|\hat{\calA}_0|^2 = \Omega^{-2} |\calA_0|^2.
	\label{A0NormSc}
\end{equation}
Indeed, the leading-order stress-energy tensor is preserved up to scale:
\begin{equation}
	\hat{\mathcal{T}}^0_{ab} = \Omega^{-2} \mathcal{T}^0_{ab} .
	\label{T0xForm}
\end{equation}
A material observer with 4-velocity $u^a$ may be used to, e.g., compute an averaged momentum flux $\langle p_a \rangle \equiv - \langle T_{ab} \rangle u^b$ or an averaged energy density $\epsilon \equiv \langle T_{ab} \rangle u^a u^b$. Transformation rules for these quantities require that observers be identified in both metrics, but if $u^a$ remains timelike with respect to $\hat{g}_{ab}$, a transformed 4-velocity $\hat{u}^a$ can be defined by \eqref{uXform}. Then, at leading order, it is clear from \eqref{T0xForm} that $\langle \hat{p}_a \rangle$ and $\hat{\epsilon}$ are both local rescalings of their background counterparts.

Related to $\mathcal{T}^0_{ab}$ is the divergence-free current $J_0^a = |\calA_0|^2 g^{ab} k_b$ which is associated with the second law of geometric optics stated at the end of Sect. \ref{Sect:Review}. That this is conserved may be interpreted as the fact that intensity variations in geometric optics are related to changing cross-sectional areas of ray bundles. Here, \eqref{kHat} and \eqref{mapScalar} imply that
\begin{equation}
	\hat{J}^a_0 = \frac{ J^a_0 }{\Omega^4 (1 + \frac{1}{2} k \cdot w ) } .
	\label{J0sc} 
\end{equation}
The coefficient in the denominator here may be identified as the proportionality factor \eqref{epsHat} which relates volume elements associated with the background and transformed metrics, implying that the 3-form dual to $J_0^a$ is invariant with respect to all metric transformations considered here:
\begin{equation}
	\hat{\epsilon}_{abcd} \hat{J}^d_0 = \epsilon_{abcd} J^d_0.
	\label{J0dualsc}
\end{equation}
Integral forms of the conservation laws associated with $J_0^a$ and $\hat{J}_0^a$ are thus independent of $\Omega$, $w_a$, and $Y$.

\subsection{Geometric optics for electromagnetic fields}
\label{Sect:GeoEM}

The geometric-optics description of an electromagnetic vector potential in Lorenz gauge is furnished by an eikonal $\varphi$ together with a covector amplitude $\calA^0_a$ which factorizes into a scalar component $\calA_0$ and a parallel-propagated polarization covector $e_a$; cf. \eqref{A0s}. As transformation laws for $\varphi$ and $\calA_0$ have already been established, all that remains to understand metric invariance for electromagnetic fields in geometric optics is to find an appropriate transformation $e_a \mapsto \hat{e}_a$.

To review, it was shown in Paper I that if the null basis \eqref{tetrad} is parallel transported along the optical rays, the polarization covector may be expanded as
\begin{equation}
	e_a = e_+ m_a + e_- \bar{m}_a + \chi k_a,
	\label{eExpand}
\end{equation}
where $e_\pm$ and $\chi$ are constant along rays. We now assume that at least one of the $e_\pm$ is nonzero along a given ray, in which case the term involving $\chi$ is pure gauge at leading order. A ray is then said to be circularly polarized if $e_a$ is null or linearly polarized if $\epsilon_{abcd} g^{be} g^{cf} g^{dg} e_e \bar{e}_f k_g = 0$, conditions which are equivalent to
\begin{equation}
	e_+ e_- = 0 \quad \mbox{(circ. pol.)}, \qquad |e_+|^2 = |e_-|^2 \quad \mbox{(lin. pol.)}.
	\label{pol}
\end{equation}
More generally,
\begin{equation}
	\left| \frac{ |e_+|^2 - |e_-|^2 }{ |e_+|^2 + |e_-|^2} \right| \in [0,1]
	\label{circ}
\end{equation}
may be used to characterize the ``circularity'' of a ray's polarization state.

An $e_a$ compatible with $g_{ab}$ may now be mapped into an $\hat{e}_a$ compatible with $\hat{g}_{ab}$ by making use of the transformation $m_a \mapsto \hat{m}_a$ which is given by \eqref{mTilde}. The resulting $\hat{m}_a$ is normalized with respect to $\hat{g}_{ab}$ in the sense that $\hat{g}^{ab} k_a \hat{m}_b = \hat{g}^{ab} \hat{m}_a \hat{m}_b = 0$ and $\hat{g}^{ab} \hat{m}_a \hat{\bar{m}}_b = 1$. Besides the explicit appearance of the metric deformation functions $\Omega$ and $Y$, the definition of $\hat{m}_a$ also involves a complex scalar $l$ and a real scalar $\theta$, the latter of which are to be chosen such that parallel propagation of $m_a$ with respect to $g_{ab}$ implies parallel propagation of $\hat{m}_a$ with respect to $\hat{g}_{ab}$. Having made these choices, it is natural to transform polarization tensors using
\begin{equation}
	e_a  \mapsto \hat{e}_a \equiv e_+ \hat{m}_a + e_- \hat{\bar{m}}_a + \chi k_a.
	\label{hate}
\end{equation}
Doing so preserves magnitudes in the sense that $g^{ab} e_a \bar{e}_b = \hat{g}^{ab} \hat{e}_a \hat{\bar{e}}_b$, and also gross features such as the polarization circularity defined by \eqref{circ}.

A more detailed comparison between $e_a$ and $\hat{e}_a$ arises by expanding the latter explicitly in terms of the background tetrad. Eqns. \eqref{hate} and \eqref{mTilde} imply that
\begin{align}
	\hat{e}_a = \frac{ \Omega }{ (1 - \frac{1}{4} |Y|^2 )^{\frac{1}{2} } } \Big\{  \left[ ( e_+ e^{i \theta} + \tfrac{1}{2} \bar{Y} e_- e^{-i \theta} ) m_a + ( e_- e^{-i \theta} + \tfrac{1}{2} Y e_+ e^{i \theta} ) \bar{m}_a \right] 
	\nonumber
	\\
	~ + \left[\Omega^{-1} \chi(1 - \tfrac{1}{4} |Y|^2 )^{ \frac{1}{2} }  + e_+ l e^{i \theta} + e_- \bar{l} e^{-i \theta}  \right]k_a \Big\},
	\label{hateExp}
\end{align}
which depends on the metric deformation functions locally via the explicit appearance of $\Omega$ and $Y$, and also nonlocally via the deformation-dependent transport equations satisfied by $l$ and $\theta$. Regardless, it is clear that except in special cases, $\hat{e}_a$ is not equal---or even proportional---to $e_a$. In this sense, the mapping from $\calA^0_a = \calA_0 e_a$ to $\hat{\calA}^0_a = (\calA_0/\Omega) \hat{e}_a$ is not local and algebraic for the full space of metric transformations generated by \eqref{gTilde}; observables which are sensitive to fine details of a field's polarization state might not be preserved.

Such details \textit{are} however preserved when $Y=0$. It follows from \eqref{theta} that in these cases, it is possible to set $\theta = 0$, which implies that
\begin{equation}
	\hat{e}_a = \Omega \left\{ e_a + \left[ (\Omega^{-1}-1) \chi +  e_+ l + e_- \bar{l} \right] k_a \right\}
	\label{eXform}
\end{equation}
for the restricted class of metric transformations
\begin{equation}
	g_{ab} \mapsto \hat{g}_{ab} = \Omega^2 (g_{ab} + k_{(a} w_{b)} ). 
	\label{gTildeSimp}
\end{equation}
Crucially, the inhomogeneous term which is proportional to $k_a$ in \eqref{eXform} is pure gauge; it does not affect the leading-order field strength [cf. \eqref{f0HatSimp} below].

It follows from \eqref{A0s}, \eqref{mapScalar}, \eqref{eExpand}, and \eqref{hate} that for metric transformations generated by arbitrary $\Omega$, $w_a$, and $Y$, the $n=0$ optical fields associated with electromagnetic waves transform via
\begin{align}
	O_0 = \{ g_{ab} ; &\varphi, \calA_0 (e_+ m_a + e_- \bar{m}_a + \chi k_a) \} \mapsto 
	\nonumber
	\\
	&\hat{O}_0 = \{ \hat{g}_{ab} ; \varphi, \Omega^{-1} \calA_0  ( e_+ \hat{m}_a + e_- \hat{\bar{m}}_a + \chi k_a) \},
		\label{O0ex}
\end{align}
where the $\hat{m}_a$ appearing here is given by \eqref{mTilde}. If $Y=0$, this simplifies to
\begin{equation}
	\{ g_{ab} ; \varphi, \calA^0_a \} \mapsto 
	\{ \hat{g}_{ab} ; \varphi, \calA^0_a + (\ldots) k_a  \}.
	\label{O0ex2}
\end{equation}

The physical consequences of these identifications may be better understood by computing the corresponding (gauge invariant) field strengths. Although it is the vector potential $\psi_a$ which is expanded here in a high-frequency limit, it is the (complexified) field strength $f_{ab} = 2 \nabla_{[a} \psi_{b]}$ which is more directly connected to observations. Following Paper I, it is convenient to introduce coefficients $\mathcal{F}_{ab}^n$ which characterize this via
\begin{equation}
	f_{ab}(x;\omega) = - 2 i  \omega e^{i \omega \varphi} \sum_{n=0}^\infty \omega^{-n} \mathcal{F}^n_{ab}(x),
	\label{fExpand}
\end{equation}
the first of which is $\mathcal{F}_{ab}^0 = k_{[a} \calA_{b]}^0$. For any of the transformations considered here, including those in which $Y \neq 0$,
\begin{equation}
	\mathcal{F}_{ab}^0 = \calA_0 k_{[a} ( e_+ m_{b]} + e_-  \bar{m}_{b]} )
	\label{Fcal0}
\end{equation}
and
\begin{align}
	\hat{\mathcal{F}}_{ab}^0 &= \Omega^{-1} \calA_0 k_{[a} ( e_+ \hat{m}_{b]} + e_-  \hat{\bar{m}}_{b]} )
	\nonumber
	\\
	& = \frac{ \calA_0 }{ (1 - \frac{1}{4} |Y|^2 )^{\frac{1}{2}} } k_{[a} \left[ ( e_+ e^{i \theta} + \tfrac{1}{2} \bar{Y} e_- e^{-i \theta} ) m_{b]} + (e_- e^{-i \theta} + \tfrac{1}{2} Y e_+ e^{i \theta}) \bar{m}_{b]} \right] .
	\label{f0Hat}
\end{align}
As already alluded to, the tensorial structure of the electromagnetic field is not necessarily preserved. However, \eqref{O0ex2} implies that if $Y=0$ and $\theta$ is chosen to vanish,
\begin{equation}
	\hat{\mathcal{F}}_{ab}^0 = \mathcal{F}_{ab}^0.
	\label{f0HatSimp}
\end{equation}
This is related to the exact result \cite{HarteEMKS} that metric transformations with the form \eqref{gTildeSimp} preserve null electromagnetic fields with principal null direction $k_a$. 

Despite the complexity of \eqref{f0Hat} as compared with its $Y=0$ specialization \eqref{f0HatSimp}, a number of electromagnetic observables obey simple transformation rules even when $Y \neq 0$. One example is the aforementioned polarization circularity. Another is the averaged stress-energy tensor: Again expanding $\langle T_{ab} \rangle$ via \eqref{Texpand}, it is shown in Paper I that the leading-order contribution is controlled by $\mathcal{T}^0_{ab} = |\calA_0|^2 k_a k_b$, where $|\calA_0|^2 \equiv g^{ab} \calA^0_a \bar{\calA}^0_b$. However, it follows from \eqref{O0ex} that even if $Y \neq 0$,
\begin{equation}
	|\hat{\calA}_0|^2 = \Omega^{-2} |\calA_0|^2.
	\label{A0NormEM}
\end{equation}
This is identical to its scalar counterpart \eqref{A0NormSc}. Indeed, all of the electromagnetic $\mathcal{T}_{ab}^0$ transforms identically to its scalar counterpart \eqref{T0xForm}. Similar comments also apply for the electromagnetic conserved current $J_0^a =|\calA_0|^2 g^{ab} k_b$ and its dual; both \eqref{J0sc} and \eqref{J0dualsc} remain valid for electromagnetic fields.

Observables associated with charged-particle motion behave somewhat differently from their scalar counterparts. If a test particle with charge $q$ and background 4-velocity $u^a$  is subject to a high-frequency electromagnetic field, and if $u^a$ is timelike with respect to both $g_{ab}$ and $\hat{g}_{ab}$, forces $f_a = q f_{ab} u^b$ may be compared in both geometries by using \eqref{uXform} to transform 4-velocities and \eqref{fExpand} and \eqref{f0Hat} to transform $f_{ab}$. For a general $\hat{g}_{ab}$, it is clear that $f_{a}$ is not necessarily proportional to $\hat{f}_{a} = q \hat{f}_{ab} \hat{u}^b$ at leading order---implying that geometries might be distinguished by measuring the forces which act on material bodies. However, it is really only $Y$ which is so distinguished. It follows from \eqref{f0HatSimp} that forces are at most rescaled when $Y=0$, and this rescaling can be attributed entirely to the difference in proper times associated with the two metrics. It is however important to distinguish between forces and accelerations: Even if $Y = 0$ so $f_a$ is proportional to $\hat{f}_a$, the acceleration $g^{ab} f_b/m$ is not necessarily proportional to $\hat{g}^{ab} \hat{f}_b/m$.

\subsection{Geometric optics for gravitational waves}
\label{Sect:GOgrav}

Using \eqref{A0s}, the leading-order gravitational amplitude $\calA^0_{ab}$ may be decomposed into a scalar component $\calA_0$ and a parallel-transported polarization tensor $e_{ab} = e_{(ab)}$. Again assuming that $m_a$ is parallel transported along the optical rays, it was shown in Paper I that all polarization tensors can be written as
\begin{equation}
	e_{ab} = e_+ m_a m_b + e_- \bar{m}_a \bar{m}_b + k_{(a} \chi_{b)},
	\label{eExpandGrav}
\end{equation}
where $e_\pm$ are constants along each ray and $\chi_a$ is parallel transported. If at least one of the $e_\pm$ is nonzero, the term involving $\chi_a$ is pure gauge at leading order. Linear and circular polarization may then be defined in the same way as in electromagnetism; cf. \eqref{pol} and \eqref{circ}.

Following the electromagnetic discussion above in which the background polarization \eqref{eExpand} is replaced with \eqref{hate}, it is natural to let
\begin{equation}
	e_{ab} \mapsto \hat{e}_{ab} = e_+ \hat{m}_a \hat{m}_b + e_- \hat{\bar{m}}_a \hat{\bar{m}}_b + k_{(a} \hat{\chi}_{b)}
	\label{eHatGrav}
\end{equation}
for gravitational waves propagating on a background $\hat{g}_{ab}$, where $\hat{m}_a$ is again given by \eqref{mTilde}.  We leave $\hat{\chi}_a$ unspecified except to say that it must be parallel transported with respect to $\hat{g}_{ab}$. Recalling \eqref{A0s} and \eqref{mapScalar}, the zeroth-order optical fields for gravitational waves thus transform via
\begin{align}
	O_0 = \{ g_{ab} ;& \varphi, \calA_0 (e_+ m_a m_b + e_- \bar{m}_a \bar{m}_b + k_{(a} \chi_{b)}) \} \mapsto 
	\nonumber
	\\
	&\hat{O}_0 = \{ \hat{g}_{ab} ; \varphi, \Omega^{-1} \calA_0 (e_+ \hat{m}_a \hat{m}_b + e_- \hat{\bar{m}}_a \hat{\bar{m}}_b + k_{(a} \hat{\chi}_{b)} \}.
	\label{O0grav}
\end{align}
As in the electromagnetic case, $\calA^0_{ab}$ and $\hat{\calA}^0_{ab}$ are not necessarily proportional when  $Y \neq 0$. However, if $Y=0$ and $\theta$ is again chosen to vanish, \eqref{mTilde} implies that
\begin{equation}
	 \{ g_{ab} ; \varphi, \calA^0_{ab} \} \mapsto 
	 \{ \hat{g}_{ab} ; \varphi, \Omega \calA^0_{ab} + k_{(a} (\ldots)_{b)} \}.
	 	\label{O0gravSimp}
\end{equation}
The omitted term on the right-hand side here is pure gauge at leading order, implying that the physical aspects of gravitational waves transform very simply for all metric transformations with the form \eqref{gTildeSimp}.

Recall that the amplitude $\calA^0_{ab}$ appears in an expansion of $\psi_{ab}$, which is a linearized metric perturbation in Lorenz gauge. However, it is somewhat more physical to consider linearized perturbations of the Riemann tensor. The complexified version of such a perturbation (with all indices down) is denoted by $\delta r_{abcd}$, and was expanded in Paper I as 
\begin{equation}
	\delta r_{abcd}(x;\omega) = - 2 \omega^2 e^{i \omega \varphi} \sum_{n=0}^\infty \omega^{-n} \mathcal{R}^n_{abcd}(x),
	\label{RiemExpand}
\end{equation}
where the leading-order perturbation is controlled by $\mathcal{R}^0_{abcd} = k_{[a} \calA^0_{b][c} k_{d]}$. It follows from \eqref{O0grav} and \eqref{mTilde} that $\mathcal{R}^0_{abcd}$ is not necessarily proportional to its hatted counterpart $\hat{\mathcal{R}}^0_{abcd}$, although
\begin{equation}
	\hat{\mathcal{R}}^0_{abcd} = \Omega \mathcal{R}^0_{abcd}
	\label{RiemXform} 
\end{equation}
when $Y = 0$ and $\theta$ is chosen to vanish. A gravitational wave curvature perturbation which is known in one background may therefore be used to determine curvature perturbations in a family of backgrounds which differ by the five free functions associated with arbitrary choices of $\Omega$ and $w_a$. Despite the extensive use of conformal transformations in the literature, we are not aware of \eqref{RiemXform} being noted previously even in the purely-conformal case for which $w_a = 0$. 

As in the electromagnetic case, observables which are indifferent to the fine details of a gravitational wave's polarization state obey simple transformation rules even when $Y \neq 0$. For example, the circularity \eqref{circ} of the polarization state does not depend on $\Omega$, $w_a$, or $Y$. Other examples can be constructed from perturbations to the Bel-Robinson tensor $T_{abcd}$. First recall from Paper I that $\langle \delta T_{abcd} \rangle = \omega^4 \left[ \frac{1}{16} \| \calA_0 \|^2 k_a k_b k_c k_d + \mathcal{O}(\omega^{-1}) \right]$, where
\begin{equation}
	\| \calA_0 \|^2 \equiv ( g^{ac} g^{bd} - \tfrac{1}{2} g^{ab} g^{cd} ) \calA^0_{ab} \bar{\calA}_{cd}^0. 
\end{equation}
The complexity of this norm when compared with, e.g., $g^{ac} g^{bd} \calA^0_{ab} \bar{A}^0_{cd}$, arises because i) our amplitudes describe metric perturbations and not their trace-reversed counterparts, and ii) the gauge freedom has not been used to eliminate traces. Regardless, \eqref{O0grav} implies that
\begin{equation}
	\| \hat{\calA}_0 \|^2 = \Omega^{-2} \| \calA_0 \|^2,
\end{equation}	
which is reminiscent of the scalar and electromagnetic equations \eqref{A0NormSc} and \eqref{A0NormEM}. It follows that even if $Y \neq 0$, Bel-Robinson perturbations transform as
\begin{equation}
	\langle \delta \hat{T}_{abcd} \rangle = \Omega^{-2} \langle \delta T_{abcd} \rangle + \mathcal{O}(\omega^3).
	\label{BelRobinson}
\end{equation}
Certain other observables, such as those involving the relative accelerations of freely-falling test particles, can depend nontrivially on $Y$.

\section{``Inheritance'' and metric invariance beyond geometric optics}
\label{Sect:Inheritance}

The results of Sects. \ref{Sect:Rays} and \ref{Sect:Geo} imply that individual field configurations in geometric optics depend very little on the background geometry. While measurements which go beyond geometric optics can be more discerning, this is not necessarily the case: One of the interesting results of Paper I is that many corrections to geometric optics can be locally\footnote{It follows immediately from the hierarchical structure of the transport equations \eqref{xPort} that solutions in geometric optics determine their own corrections. However, this kind of dependence is nonlocal in general. The nontrivial result is that it can sometimes be localized.} written in terms of the observables of geometric optics itself, and in some of these cases, the metric invariance which arises at leading order is effectively inherited by observables at higher orders. These are first cases we discuss.

\subsection{Inheritance for scalar fields}

It was shown in Paper I that for scalar fields, it can be convenient to introduce\footnote{The quantities denoted here by $\varphi^\mathrm{cor}$ and $k^\mathrm{cor}_a$ were written as $\hat{\varphi}$ and $\hat{k}_a$ in Paper I. This notation has been changed in order not to conflict with the present usage of hatted quantities as those associated with certain transformed metrics.} a corrected, frequency-dependent eikonal $\varphi^\mathrm{cor}$ and its associated wavevector $k^\mathrm{cor}_a$ via
\begin{equation}
	\varphi^\mathrm{cor} \equiv \varphi + \omega^{-1} \arg \calA_0, \qquad k^\mathrm{cor}_a \equiv - \nabla_a \varphi^\mathrm{cor}.
	\label{Kcorrect}
\end{equation}
Rays tangent to $g^{ab} k^\mathrm{cor}_b$ are null through leading and subleading orders, and may be interpreted as describing the first correction to a field's propagation direction. This interpretation is motivated by noting that the averaged stress-energy tensor can be written as \cite{HarteOptics1}
\begin{equation}
	\langle T_{ab} \rangle = \frac{ \omega^2 }{ 8 \pi } \left[ |\calA_0 + \omega^{-1} \calA_1 |^2 k^\mathrm{cor}_a k^\mathrm{cor}_b + \mathcal{O}(\omega^{-2}) \right] ,
	\label{TSc1}
\end{equation}
so all observers measure local momentum densities proportional to $k^\mathrm{cor}_a$. 

Now consider a general transformation \eqref{mapScalar} of the zeroth-order optical fields, including a transformed metric with the form \eqref{gTilde}. It follows immediately from the reality of $\Omega$ that $\arg \hat{\calA}_0 = \arg \calA_0$, so the corrected eikonal is preserved:
\begin{equation}
	\hat{\varphi}^\mathrm{cor} = \varphi^\mathrm{cor}, \qquad \hat{k}^\mathrm{cor}_a = k^\mathrm{cor}_a .
	\label{Kxform}
\end{equation}
This and \eqref{TSc1} imply that
\begin{equation}
	\langle \hat{T}_{ab} \rangle \propto \langle T_{ab} \rangle + \mathcal{O}(\omega^0),
	\label{TScProp}
\end{equation}
so transformed momentum densities remain proportional to their background counterparts. However, the transformed rays tangent to $\hat{g}^{ab} k^\mathrm{cor}_b$ may differ (at subleading order) from the background rays tangent to $g^{ab} k^\mathrm{cor}_b$. Also note that the proportionality factor here is fixed in \eqref{Tsc2} below, at least for conformal Kerr-Schild tranformations in which $\hat{g}_{ab}$ takes the form \eqref{gConfKS}.

Up to scale, the leading-order transformation $O_0 \mapsto \hat{O}_0$ can additionally be used to expand $\nabla_a \psi$ through subleading order. From Paper I, first note that
\begin{equation}
	\nabla_a \psi =  - i \omega |\calA_0| e^{i \omega \varphi^\mathrm{cor}} \left[ 1 + \omega^{-1} (\calA_1/\calA_0 ) \right] \left[ k^\mathrm{cor}_a + i \omega^{-1} \nabla_a \ln |\calA_0| + \mathcal{O}(\omega^{-2}) \right] .
	\label{dpsiCorrect} 
\end{equation}
As $\calA_1$ appears here only in the overall scale, \eqref{A0scalarXform} and \eqref{Kxform} imply that $\nabla_a \hat{\psi} \propto \nabla_a \psi + \mathcal{O}(\omega^{-1})$ for arbitrary $w_a$ and $Y$ but constant $\Omega$.

\subsection{Inheritance for electromagnetic fields}
\label{Sect:inheritEM}

Electromagnetic fields are more difficult to describe than their scalar counterparts. As explained in Paper I, there may fail to be any single, broadly-applicable electromagnetic analog of the scalar wavevector defined by \eqref{Kcorrect}. Instead, there are multiple inequivalent candidates at one order beyond geometric optics \cite{spinHallRev}, and most appear to be useful only in special cases.

One possibility which does have reasonably broad applicability is to consider more than one ``wavevector'' simultaneously, namely the eigen(co)vectors of $\langle T_{bc} \rangle g^{ca}$. These are discussed in Paper I. There is exactly one such eigenvector at leading order, namely $k_a$. But if the first subleading terms in the stress-energy tensor are included as well, this single eigenvector generically splits into two: Supposing that $\calA_0$ is chosen such that $|e_+|^2 + |e_-|^2 = g^{ab} e_a \bar{e}_b = 1$, they may be written as \cite{HarteOptics1}
\begin{align}
	k^{\mathrm{cor}\pm}_a = k^\mathrm{cor}_a \pm 2 \Re (\bar{z} m_a) + |z|^2 n_a - \omega^{-1} g^{bc} \Im \big[ \bar{e}_a e_b \nabla_c \ln |\calA_0|^2 + \nabla_b ( \bar{e}_a e_c) 
	\nonumber
	\\
	 - e_b \nabla_a \bar{e}_c - ( 2 g_{ab} + k_a n_b ) g^{dg} g^{fh} n_d e_f \bar{e}_{(c} \nabla_{g)} k_h \big] + \mathcal{O}(\omega^{-3/2 }) ,
	 \label{Kpm}
\end{align}
where 
\begin{equation}
	z \equiv [ (|e_+|^2 - |e_-|^2 ) i \sigma/\omega]^{ \tfrac{1}{2} }
	\label{zDef}
\end{equation}
is $\mathcal{O}(\omega^{-1/2})$, $\sigma$ denotes the shear \eqref{sigDef} of the rays, and $k^\mathrm{cor}_a$ is given by \eqref{Kcorrect}. Both $k^{\mathrm{cor}+}_a$ and $k^{\mathrm{cor}-}_a$ are null to the relevant order, and in terms of them, the averaged stress-energy tensor is given by
\begin{equation}
	\langle T_{ab} \rangle = \frac{ \omega^2 }{ 8\pi} |\calA_0 + \omega^{-1} \calA_1|^2 \left[ (\delta^c_{(a} \delta^d_{b)} - \tfrac{1}{4} g_{ab} g^{cd} ) k^{\mathrm{cor}+}_{c} k^{\mathrm{cor}-}_{d} + \mathcal{O} (\omega^{-2}) \right].
	\label{Tem}
\end{equation}
These eigenvectors are related to the principal null directions of $F_{ab} = \Re f_{ab}$, although they are not afflicted by the rapid oscillations of those directions which arise for fields which are not circularly polarized. We now ask how the $k^{\mathrm{cor}\pm}_a$ transform when $g_{ab} \mapsto \hat{g}_{ab}$.

It is simplest to begin with linearly-polarized fields. Given \eqref{pol} and \eqref{zDef}, these correspond to cases in which $z=0$ and $k^{\mathrm{cor}+}_a = k^{\mathrm{cor}-}_a$ to the order in which we work; the wavevectors remain degenerate. Little generality is lost by setting $\chi = 0$ in \eqref{hate}, and doing so for simplicity implies that $\Im (e_a \bar{e}_b) = 0$. Eq. \eqref{Kpm} then  simplifies to
\begin{equation}
	k^{\mathrm{cor}\pm}_a = k_a + \omega^{-1} [\Im ( e_+ \nabla_a \bar{e}_+ + e_- \nabla_a \bar{e}_-) -\nabla_a \arg \calA_0 ] + \mathcal{O}(\omega^{- 3/2 }) 
	\label{kClinPol}
\end{equation}
for linearly-polarized fields. Finally, inspection of \eqref{O0ex} implies that 
\begin{equation}
	\hat{k}^{\mathrm{cor}\pm}_a = k^{\mathrm{cor}\pm}_a
	\label{kCorrectEM}
\end{equation}
for metric transformations generated by arbitrary $\Omega$, $w_a$, and $Y$. In fact, this result is not very different from its scalar counterpart \eqref{Kxform}: The splitting of $\calA^0_a = \calA_0 e_a$ into $\calA_0$ and $e_a$ is not unique, in that the latter quantities may be rescaled along each ray in such a way that their product remains unchanged. Some of this ambiguity has already been fixed by requiring that $|e_+|^2 + |e_-|^2 = 1$, although there remains a freedom to let $\calA_0 \mapsto e^{i \kappa} \calA_0$ and $e_a \mapsto e^{-i \kappa} e_a$, where $\kappa$ is real. While \eqref{kClinPol} is invariant under all such transformations, there always exists a particular choice of $\kappa$ for which the $\Im ( \ldots)$ terms in that expression vanish and $k^{\mathrm{corr} \pm}_a = k^\mathrm{cor}_a + \mathcal{O}(\omega^{-3/2})$. In this sense, linearly-polarized electromagnetic fields propagate identically to scalar fields, at least through leading and subleading orders. Nevertheless, some care may be required to identify $\kappa$, which corresponds to fixing the phase characteristics of a ``comparable'' scalar field.

In the more generic setting for which $(|e_+|^2 - |e_-|^2 ) \sigma \neq 0$, the wavectors $k_a^{\mathrm{cor}+}$ and $k_a^{\mathrm{cor}-}$ differ from one another already at $\mathcal{O}(\omega^{-1/2})$. Transformations of these differences are relatively simple at least when $Y=0$: In those cases, \eqref{Kpm}, \eqref{zDef}, \eqref{mTilde}, and \eqref{sigxForm} imply that
\begin{equation}
	\hat{k}^{\mathrm{cor} \pm}_a \propto k_a \pm \frac{ 2  \Re (\bar{z} m_a ) }{ (1 + \frac{1}{2} k \cdot w )^{ \frac{1}{2} } } + \mathcal{O}(\omega^{-1}).
	\label{kCEM}
\end{equation}
The hatted and unhatted eigenvectors therefore differ at this order (by more than an overall factor) whenever $k \cdot w \neq 0$. No such difference appears if the class of metric transformations is further narrowed to the conformal Kerr-Schild class
\begin{equation}
	g_{ab} \mapsto \hat{g}_{ab} = \Omega^2 ( g_{ab} + V k_a k_b)
	\label{gConfKS}
\end{equation}
in which $Y=0$ and $w_a = V k_a$ for some scalar $V$.

Besides aspects of the averaged stress-energy tensor associated with $k^{\mathrm{cor}\pm}_a$, it was shown in Paper I that the first nontrivial terms in high-frequency expansions for the complex Newman-Penrose scalars $\Phi_0$, $\Phi_1$, and $\Phi_2$ can also be written locally in terms of leading-order quantities, despite that $\Phi_0$ and $\Phi_1$ describe components of the electromagnetic field which are not a part of geometric optics itself. First considering $\Phi_2$, which does characterize just the geometric-optics field at leading order, it was shown in Paper I that
\begin{align}
	\Phi_2 &\equiv F_{ab} g^{ac} g^{bd} \bar{m}_c n_d
	\nonumber
	\\
	 &= - \tfrac{1}{2} i \omega |\calA_0| (e_+ e^{i \omega \varphi^\mathrm{cor} } - \bar{e}_- e^{-i \omega \varphi^\mathrm{cor}} ) + \mathcal{O}(\omega^0).
	 \label{Phi2}
\end{align}
Transforming the metric using arbitrary $\Omega$, $w_a$, and $Y$, it follows from \eqref{A0scalarXform}, \eqref{hate}, and \eqref{Kxform} that
\begin{align}
	\hat{\Phi}_2 = \Omega^{-1} \Phi_2 + \mathcal{O}(\omega^0).
\end{align}
The deformation functions $w_a$ and $Y$ do not affect $\Phi_2$ at leading order.

Very similar arguments may be used to characterize that portion of the subleading electromagnetic field which is described by the first nonzero term in an expansion for $\Phi_0$. First recall from Paper I that
\begin{equation}
	\Phi_0 = - \tfrac{1}{2} \sigma |\calA_0| (e_+ e^{i \omega \varphi^\mathrm{cor}} + \bar{e}_- e^{-i \omega \varphi^\mathrm{cor}} ) + \mathcal{O}(\omega^{-1}).
	\label{Phi0}
\end{equation}
If the rays are shear-free, $\Phi_0$ is suppressed by at least two powers of $\omega^{-1}$ with respect to $\Phi_2$. If $\sigma \neq 0$, it is suppressed by only a single power. Assuming the latter, the transformations \eqref{O0ex} and \eqref{sigxForm} for the optical fields and the shear imply that
\begin{align}
	\hat{\Phi}_0 = \Omega^{-3} e^{2i\theta}  \left[ \frac{ 1 - \frac{1}{4} (Y^2/\sigma) ( \bar{\sigma} + \mathcal{L}_k \ln Y^2) }{ (1-\frac{1}{4} |Y|^2 ) (1 + \frac{1}{2} k \cdot w ) } \right] \Phi_0 + \mathcal{O}(\omega^{-1}),
\end{align}
where $\theta$ is a solution to the transport equation \eqref{theta}. For conformal Kerr-Schild transformations \eqref{gConfKS}, $Y = k \cdot w = 0$ and $\theta$ may be chosen to vanish, so $\hat{\Phi}_0 = \Omega^{-3} \Phi_0 + \mathcal{O}(\omega^{-1})$.

The most complicated of the Newman-Penrose scalars which describe a real electromagnetic field is $\Phi_1$. From Paper I,
\begin{align}
	\Phi_1 = - \tfrac{1}{2} m^a |\calA_0| \big\{ e_+ \big[ \bar{m}^b \nabla_a m_b + \nabla_a \ln ( e_+ \calA_0) \big] e^{i \omega \varphi^\mathrm{cor} } + \bar{e}_-  \big[ \bar{m}^b \nabla_a m_b 
	\nonumber
	\\
	~ + \nabla_a \ln (\bar{e}_- \bar{\calA}_0 ) \big]e^{-i \omega \varphi^\mathrm{cor} } \big\} + \mathcal{O}(\omega^{-1}) . 
	\label{Phi1}
\end{align}
This is tedious to transform in general due to the terms involving $m^a \bar{m}^b \nabla_a m_b$, so we restrict for simplicity to the conformal Kerr-Schild transformations. Then,
\begin{align}
	\hat{\Phi}_1 = \Omega^{-2} ( \Phi_1 + \bar{l} \Phi_0 ) + \mathcal{O}(\omega^{-1}),
	\label{Phi1Hat}
\end{align}
where $l$ appears in the definition \eqref{mTilde} of $\hat{m}_a$ and satisfies the transport equation \eqref{lEvolve}. Note that the inhomogeneous $\Phi_0$ term which appears here is insignificant when $\sigma = 0$, as $\Phi_0$ then vanishes at order $\omega^0$. The inhomogeneous term can also disappear when $\Omega$ is constant, in which case $l$ can be chosen to vanish.

\subsection{Inheritance for gravitational waves}
\label{Sect:inheritGW}

Like electromagnetic waves, gravitational waves may be associated with principal null directions. Although there are four such directions in general, they all degenerate to $k_a$ in the geometric-optics limit. They appear to have been computed beyond this only for circularly-polarized waves in which\footnote{The opposite helicity, described by $e_+ = 0$ and $e_- = 1$, follows by swapping $m_a$ and $\bar{m}_a$ wherever they appear.} $e_+ = 1$ and $e_- =0  $, and in that case, they may be expanded as \cite{HarteOptics1}
\begin{equation}
	k^{\mathrm{cor} \pm\pm}_a = k_a \pm 2 (3 \pm \sqrt{6})^{ \frac{1}{2} } \Re ( \bar{z}  m_a ) + \mathcal{O}(\omega^{-1}),
	\label{KpmGrav}
\end{equation}
where $z= ( i \sigma/\omega)^{\frac{1}{2}}$. If the metric is now transformed using arbitrary $\Omega$ and $w_a$ but vanishing $Y$, it follows from \eqref{mTilde} and \eqref{sigxForm} that
\begin{equation}
	\hat{k}^{\mathrm{cor}\pm\pm} \propto k_a \pm \frac{ 2 (3 \pm \sqrt{6})^{ \frac{1}{2} } \Re (\bar{z} m_a ) }{ (1 + \frac{1}{2} k \cdot w)^{ \frac{1}{2} } } + \mathcal{O}(\omega^{-1}).
\end{equation}
This is closely analogous to its electromagnetic equivalent \eqref{kCEM}; in both cases, wavevectors are preserved to the given order when $k \cdot w = 0$ or $\sigma = 0$. 

It is also possible to consider transformation laws for the Newman-Penrose scalars $\delta \Psi_0, \ldots, \delta \Psi_4$ associated with the perturbed Weyl tensor. The geometric-optics curvature is fully determined by $\delta \Psi_4$, and although the other scalars characterize aspects of the curvature which go beyond geometric optics, they too can be locally written using only $O_0$. Beginning with $\delta \Psi_4$, it was shown in Paper I that to leading order, 
\begin{align}
	\delta \Psi_4 &\equiv \delta C_{abcd} k^a m^b k^c m^d 
	\nonumber
	\\
	&= \tfrac{1}{4} \omega^2 |\calA_0| ( e_+ e^{i \omega \varphi^\mathrm{cor} } + \bar{e}_- e^{-i \omega \varphi^\mathrm{cor}  } ) + \mathcal{O}(\omega).
	\label{Psi4}
\end{align}
This may be transformed using \eqref{O0grav} and \eqref{Kxform}, from which it follows that for arbitrary $\Omega$, $w_a$, and $Y$,
\begin{equation}
	\delta \hat{\Psi}_4 = \Omega^{-1} \delta \Psi_4 + \mathcal{O}(\omega).
\end{equation}
The next simplest Newman-Penrose scalar can be written as \cite{HarteOptics1}
\begin{equation}
	\delta \Psi_0 = - 3(\sigma/\omega)^2 \delta \Psi_4 + \mathcal{O}(\omega^{-1}),
	\label{Psi0}
\end{equation}
and using the transformation \eqref{sigxForm} for the shear, it follows that
\begin{equation}
	\delta \hat{\Psi}_0 = \Omega^{-5} e^{4 i \theta } \left[ \frac{  1 - \frac{1}{4} (Y^2/\sigma) (\bar{\sigma} + \mathcal{L}_k \ln Y^2 )}{ (1-\frac{1}{4}|Y|^2) (1 + \frac{1}{2} k \cdot w)} \right]^2 \delta \Psi_0 + \mathcal{O}(\omega^{-1}). 
\end{equation}
For conformal Kerr-Schild transformations in which $\theta$ is chosen to vanish, this simplifies to $\delta \hat{\Psi}_0 = \Omega^{-5} \delta \Psi_0 + \mathcal{O}(\omega^{-1})$. 

The Newman-Penrose scalar $\delta \Psi_2$ is easily transformed as well. To leading nontrivial order, its background expression was found in Paper I to be given by
\begin{equation}
	\delta \Psi_2 = \tfrac{1}{2} \omega \sigma \bar{m}^a \bar{m}^b \Im (\calA^0_{ab} e^{i \omega \varphi}) + \mathcal{O}(\omega^0),
	\label{Psi2}
\end{equation}
which is equivalent to
\begin{equation}
	\delta \Psi_2 = - \tfrac{1}{4} i \omega \sigma |\calA_0| ( e_+ e^{i \omega \varphi^\mathrm{cor} } - \bar{e}_- e^{-i \omega \varphi^\mathrm{cor} } ) + \mathcal{O}( \omega^{0} ) .
\end{equation}
Use of \eqref{O0grav} and \eqref{Kxform} shows that transforming the metric merely rescales this. For arbitrary $\Omega$, $w_a$, and $Y$,
\begin{align}
	\delta \hat{\Psi}_2 = \Omega^{-3} e^{2i \theta} \left[ \frac{  1 - \frac{1}{4} (Y^2/\sigma) (\bar{\sigma} + \mathcal{L}_k \ln Y^2 )}{ (1-\frac{1}{4}|Y|^2) (1 + \frac{1}{2} k \cdot w)} \right] \delta \Psi_2 + \mathcal{O}(\omega^0).
\end{align}
Restricting to conformal Kerr-Schild transformations in which $\theta$ is again chosen to vanish, this simplifies to $\delta \hat{\Psi}_2 = \Omega^{-3} \delta \Psi_2 + \mathcal{O}(\omega^0)$. 

The most complicated of the Newman-Penrose scalars considered here is $\delta \Psi_3$, and an expression for this which is found in Paper I may be rewritten as
\begin{align}
	\delta \Psi_3 = - \tfrac{ 1 }{ 4 } i \omega m^a |\calA_0| \big\{ e_+ \big[ 2 \bar{m}^b \nabla_a m_b + \nabla_a \ln (e_+ \calA_0) - n^b \nabla_b k_a ] e^{i \omega \varphi^\mathrm{cor} } 
	\nonumber
	\\
	~ -  \bar{e}_- \big[ 2 \bar{m}^b \nabla_a m_b + \nabla_a \ln (\bar{e}_- \bar{\calA}_0) - n^b \nabla_b k_a ] e^{-i \omega \varphi^\mathrm{cor} } \big\} + \mathcal{O}(\omega^0).
	\label{Psi3}
\end{align}
We do not discuss how this behaves under a fully general metric transformation, but instead restrict only to the conformal Kerr-Schild case \eqref{gConfKS}. Then,
\begin{equation}
	\delta \hat{\Psi}_3 = \Omega^{-2} ( \delta \Psi_3 + \bar{l} \delta \Psi_2 ) + \mathcal{O}(\omega^0), 
\end{equation}
where again, $l$ satisfies \eqref{lEvolve}. This is very similar to the electromagnetic transformation \eqref{Phi1Hat} for $\Phi_1$. The inhomogeneous term involving $\delta \Psi_2$ in this case disappears if $\sigma = 0$,  or if $\Omega$ is constant so $l$ may be chosen to vanish.

A discussion of $\delta \Psi_1$ would be more complicated and is omitted here. Also note that the results in this subsection should be understood as meaningful only when $g_{ab}$ and $\hat{g}_{ab}$ both satisfy the vacuum Einstein equation. Otherwise, the formalism used here to describe gravitational wave propagation cannot be trusted beyond the geometric optics regime; see the discussion in Sect. \ref{Sect:SubleadingGW} below.

\section{Subleading amplitudes}
\label{Sect:BeyondGO}

Metric transformations generated by $\Omega$, $w_a$, and $Y$ do little to affect many of the quantities of interest in geometric optics. They also have relatively little effect on many of the inherited quantities discussed in Sect. \ref{Sect:Inheritance}. However, this simplicity is expected to break down eventually, at least if---as assumed here---the metric transformations themselves do not depend on $\omega$. Non-conformal metric transformations which preserve a given field must be adapted to it in some way, and the adaptation of our transformations to $k_a$ is guaranteed to be physically relevant only in the $\omega \to \infty$ limit. This may be seen more precisely for Maxwell fields, where exact metric-invariance results are known to involve transformations adapted to a field's principal null directions \cite{HarteEMKS}. These directions typically vary with $\omega$, so broadly-applicable invariance results must involve metric transformations which also depend on $\omega$.

While it would be interesting to construct $\omega$-dependent metric transformations---especially for the scalar and gravitational cases in which no exact results are known---it might still be asked when the $\omega$-independent transformations considered above cease to be relevant. The first significant issues arise already for the subleading amplitudes $\calA^1_B$, and we now turn our attention to these. To simplify the analysis, we eliminate five of the seven deformation functions in \eqref{gTilde} by setting $Y=0$ and $w_a = V k_a$, which leaves the conformal Kerr-Schild metric transformations given by \eqref{gConfKS}. Roughly speaking, it is found that although the $\calA^1_B$ are not necessarily preserved even in this restricted context, differences can be interpreted essentially in terms of nontrivial phase shifts. 

\subsection{Subleading scalar amplitudes}

For high-frequency scalar fields, a subleading background amplitude $\calA_1$ must satisfy the transport equation $L \calA_1 = - i \mathcal{D} \calA_0 = - i (\Box - \xi R - \mu^2) \calA_0$. If the background metric is transformed using \eqref{gConfKS}, the associated transformation $L \mapsto \hat{L}$ of the transport operator is given by \eqref{Ltilde} with $k \cdot w =0$. Understanding how the source in the transport equation transforms additionally requires that we determine how $\hat{\mathcal{D}} \hat{\calA}_0 = \hat{\mathcal{D}} (\calA_0/\Omega)$ differs from $\mathcal{D} \calA_0$. First addressing the wave operator portion of $\mathcal{D}$, a direct calculation shows that
\begin{align}
	\Omega^3 \hat{\Box} (\calA_0/\Omega) = \Box \calA_0 + \big\{  \Omega^{-1} \big[ \nabla_a ( V k^a \mathcal{L}_k \Omega) - \Box  \Omega \big]
	 + \tfrac{1}{4} L (V \nabla \cdot k ) \big\} \calA_0.
\end{align}
It is also necessary to compute the transformed Ricci scalar
\begin{align}
	\hat{R} = \Omega^{-2} \big\{ R + L ( V \nabla \cdot k) + V \nabla^a k^b \nabla_a k_b + \mathcal{L}_k^2 V + 6 \Omega^{-1} [\nabla_a ( V k^a \mathcal{L}_k \Omega ) - \Box \Omega ] \big\}
\end{align}
at least when the curvature coupling $\xi$ is nonzero. Combining these expressions,
\begin{align}
	\Omega^3 \hat{\mathcal{D}}(\calA_0/\Omega) = \mathcal{D} \calA_0 + \big\{ \tfrac{1}{4} (1-4 \xi) L (V \nabla \cdot k )  - \xi ( V \nabla^a k^b \nabla_a k_b + \mathcal{L}_k^2 V )
	\nonumber
	\\
	~   + (1-6 \xi) \Omega^{-1} [ \nabla_a ( V k^a \mathcal{L}_k \Omega)  - \Box \Omega ]  - \mu^2 (\Omega^2-1) \big\} \calA_0 .
	\label{xFormDSc}
\end{align}
The conformal factor disappears here in the massless case for which $\xi = 1/6$, which is expected given that this is the value of the curvature coupling which is known \cite{Wald} to make the massless Klein-Gordon equation conformally invariant. 

With regards to determining the source in the transport equation for $\hat{\calA}_1$, the important point in this calculation is that $\Omega^3 \hat{\mathcal{D}}(\calA_0/\Omega) - \mathcal{D} \calA_0$ must be proportional to $\calA_0$.  Given \eqref{xPort} and \eqref{Ltilde}, it follows from \eqref{xFormDSc} that one solution for a transformed subleading amplitude is
\begin{equation}
	\hat{\calA}_1 = \Omega^{-1} ( \calA_1 - i \vartheta \calA_0),
	\label{A1Hat}
\end{equation}
where 
\begin{align}
	2 \mathcal{L}_k \vartheta = \tfrac{1}{4} (1-4 \xi) L (V \nabla \cdot k )  - \xi ( V \nabla^a k^b \nabla_a k_b + \mathcal{L}_k^2 V )+ (1-6 \xi) \Omega^{-1}
	\nonumber
	\\
	~  \times  [ \nabla_a ( V k^a \mathcal{L}_k \Omega) - \Box \Omega ]  - \mu^2 (\Omega^2-1).
	\label{varthetaEvolve}
\end{align}
That the right-hand side here is real means that $\vartheta$ can itself be chosen to be real, and in that case, it is suggestive to substitute \eqref{A0scalarXform} and \eqref{A1Hat} into \eqref{WKB} to show that the field reduces to
\begin{equation}
	\hat{\psi} = \Omega^{-1} e^{-i \vartheta/\omega} \psi + \mathcal{O}(\omega^{-2}),
	\label{psiPhase}
\end{equation}
through leading and subleading orders. The ratio $-\vartheta/\omega$ may thus be viewed as a phase shift. It is reminiscent of the shifts discussed in Sect. 5.2 of Paper I, where scalar fields associated with differing values of the field mass or curvature coupling are related in a fixed metric. In the case considered here, the metric is instead varied while $\mu$ and $\xi$ are held fixed.

Although some effort may be required to compute $\vartheta$, there are relatively few situations in which it is necessary; most observations are not sensitive to phase shifts. For example, $\vartheta$ does not affect anything constructed from $\langle T_{ab} \rangle$. First noting
\begin{equation}
	|\hat{\calA}_0 + \omega^{-1} \hat{\calA}_1|^2 = \Omega^{-2} |\calA_0 + \omega^{-1} \calA_1|^2 + \mathcal{O}(\omega^{-2}) ,
	\label{AMagSc}
\end{equation}
it follows from \eqref{TSc1} and \eqref{Kxform} that
\begin{equation}
	\langle \hat{T}_{ab} \rangle = \Omega^{-2} \langle T_{ab} \rangle + \mathcal{O}(\omega^{0}).
	\label{Tsc2}
\end{equation}
This generalizes \eqref{T0xForm} beyond geometric optics, although only for metrics related by conformal Kerr-Schild transformations.

\subsection{Subleading electromagnetic amplitudes}

Understanding how subleading amplitudes transform for electromagnetic fields is more complicated than for scalar fields, because i) the extra index in $\mathcal{D} \calA^0_a = \Box \calA^0_a - R_{ab} g^{bc} \calA^0_c$ makes it more difficult to transform than $\mathcal{D} \calA_0$, ii) the gauge condition \eqref{gaugeEM} must be satisfied as well as the transport equation \eqref{xPort}, and iii) the leading-order electromagnetic amplitude involves an inhomogeneous component and not simply a rescaling; cf. \eqref{O0ex2}. 

As in the scalar case considered above, we restrict for simplicity to conformal Kerr-Schild transformations so $\hat{g}_{ab}$ has the form \eqref{gConfKS}. This leaves the conformal degree of freedom parametrized by $\Omega$ and the Kerr-Schild degree of freedom parametrized by $V$. However, it is well-known that Maxwell's equations, which may be written as $\rmd \star \rmd A = 0$, are conformally invariant \cite{Wald}; the exterior derivative $\rmd$ is metric independent and a  short calculation shows that the Hodge dual $\star$ of any 2-form is conformally invariant in four dimensions. Beyond this, it is known \cite{HarteEMKS} that any Maxwell solution which admits $k_a$ as a principal null direction is invariant with respect to Kerr-Schild (and other) transformations generated by $k_a$.

There are two problems with applying these results in the high-frequency context considered here. First, the transport equations we are working with produce high-frequency approximations for vector potentials in Lorenz gauge, but the Lorenz gauge condition $g^{ab} \nabla_a A_b = 0$ is not conformally invariant. This means that even for a purely conformal transformation, amplitudes satisfying our equations can acquire nontrivial gauge corrections\footnote{Gauge issues might be avoided by using the Newman-Penrose scalars discussed in Sect. \ref{Sect:inheritEM} to construct corrected field strengths. However, doing so would require knowledge of $\Phi_2$ and $\hat{\Phi}_2$ to one order beyond what was considered there, which in turn requires knowledge of at least certain components of the transformation $\calA^1_{a} \mapsto \hat{\calA}^1_a$. Isolating those components is not significantly simpler than analyzing the amplitudes in full.}. While these are in some sense irrelevant, they must be taken into account. The second problem with applying non-perturbative results here is that even though $k_a$ is a principal null direction for the leading-order geometric optics field, it does not necessarily remain a principal null direction at higher orders. A careful analysis is thus required to understand precisely how  subleading amplitudes behave under conformal Kerr-Schild transformations adapted to $k_a$.

First recall from \eqref{A0s} and \eqref{eExpand} that the background leading-order amplitude $\calA_a^0$ is controlled by $e_+$, $e_-$, and $\chi$, scalars which are constant along each ray. Although $\chi$ is pure gauge in geometric optics, it can have physical consequences beyond this \cite{HarteOptics1}. Nevertheless, there is little loss of generality in setting it to zero. Doing so, it follows from \eqref{O0ex} and \eqref{mTilde} that for any conformal Kerr-Schild transformation,
\begin{equation}
	\hat{\calA}^0_a = \calA^0_a + \calA_0 \left( e_+ l + e_- \bar{l} \right) k_a.
	\label{A0HatEM}
	\end{equation}
Here, $l$ again satisfies the transport equation \eqref{lEvolve} and $\theta$ has again been set to zero [which is allowed by \eqref{theta}]. The inhomogeneous term proportional to $k_a$ in this expression represents the leading-order gauge transformation needed to preserve the Lorenz gauge condition. If its interpretation as a gauge transformation is retained also at subleading order, one would expect $\hat{\calA}^1_a$ to differ from $\calA^1_a$ at least by $i \nabla_a [ \calA_0 (e_+ l + e_- \bar{l} )]$. Also allowing a subleading gauge transformation proportional to $k_a$ and a ``generalized phase shift'' analogous to that generated by $\vartheta$ in the scalar transformation law \eqref{A1Hat}, suppose that
\begin{equation}
	\hat{\calA}^1_a = \calA^1_a - i \vartheta_{a}{}^{b} \calA^0_b + i \nabla_a [ \calA_0 (e_+ l + e_- \bar{l} )] + \lambda k_a,
	\label{A1HatEM}
\end{equation}
where $\vartheta_{a}{}^{b}$ and $\lambda$ are to be determined by the transport and gauge conditions.

The first step is to verify that for this ansatz, the hatted form of the gauge condition \eqref{gaugeEM} is indeed satisfied. Using \eqref{kHat}, \eqref{A0HatEM}, \eqref{A1HatEM}, and \eqref{divHat}, that gauge condition reduces to
\begin{equation}
	\hat{g}^{ab} ( k_a \hat{\calA}^1_b + i \hat{\nabla}_a \hat{\calA}^0_b ) = - i \Omega^{-2} k^a \vartheta_{a}{}^{b} \calA_b^0 = 0
\end{equation}
for all conformal Kerr-Schild transformations. It is therefore necessary to assume that
\begin{equation}
	k^a \vartheta_{a}{}^{b} \calA^0_b = 0.
	\label{ktheta}
\end{equation}
Of course, it must be ensured that the subleading transport equation holds as well. Without entering into details, $\hat{L} \hat{\calA}^1_a + i \hat{\mathcal{D}} \hat{\calA}^0_a = 0$ can be computed and contracted with the various basis vectors. Contraction with $k^a$ yields no new information and  contraction with $n^a$ yields a transport equation for $\lambda$ which is omitted here. More interesting are the contractions with $m^a$ and $\bar{m}^a$, which yield the evolution equation
\begin{equation}
	k \cdot \nabla \left[ \vartheta_{a}{}^{b}- \Re ( \sigma V \bar{m}_a \bar{m}^b ) \right] = V |\sigma|^2 m_{(a} \bar{m}_{c)} g^{bc} 
	\label{varthetaEM}
\end{equation}
for $\vartheta_{a}{}^{b}$. The nontrivial trace-free component of $\vartheta_{a}{}^b$ is thus given by $\Re ( \sigma V \bar{m}_a \bar{m}^b)$, while the remaining pure-trace term must be found by integrating $V |\sigma|^2$ along rays. These results are independent of $\Omega$, as expected. Importantly for its interpretation in terms of a generalized phase shift, initial data can always be chosen such that $\vartheta_{a}{}^{b} g_{bc}$ is everywhere real and symmetric. The gauge condition \eqref{ktheta} may also be satisfied by choosing solutions in which $k^a \vartheta_{a}{}^{b} = 0$. The trivial solution $\vartheta_{a}{}^{b} = 0$ exists when the geometric-optics rays are shear-free or the transformation is purely conformal.

In order to see the effect of $\vartheta_{a}{}^{b}$ on the electromagnetic field, \eqref{A0HatEM} and \eqref{A1HatEM} may be substituted into the expansion \eqref{fExpand} to yield the complexified field strength
\begin{align}
	\hat{f}_{ab} = \exp \big( - i \delta^{c}_{[a} \vartheta_{b]}{}^{d}/\omega \big) f_{cd} + \mathcal{O}(\omega^{-1}).
	\label{fPhase}
\end{align}
This is valid through leading and subleading orders for all conformal Kerr-Schild transformations with the form \eqref{gConfKS}. Also, the reality and symmetry of $\vartheta_{ab}$ imply that the intensity prefactor which appears in $\langle T_{ab} \rangle$ does not depend on it:
\begin{align}
	|\hat{\calA}_0 + \omega^{-1} \hat{\calA}_1|^2 = \Omega^{-2} |\calA_0 + \omega^{-1} \calA_1|^2 -  2  \omega^{-1} \Omega^{-3} \nabla_a \Im \big[ \Omega (|e_+|^2 - |e_-|^2 ) 
	\nonumber
	\\
	~ \times |\calA_0|^2 l \bar{m}^a \big] +  \mathcal{O}(\omega^{-2}).
\end{align}
If a field is linearly polarized so $|e_+|^2 = |e_-|^2$, the inhomogeneous term here vanishes and it reduces to its scalar analog \eqref{AMagSc}. Moreover, \eqref{Tem} and \eqref{kCorrectEM} imply that
\begin{equation}
	\langle \hat{T}_{ab} \rangle = \Omega^{-2} \langle T_{ab} \rangle + \mathcal{O}(\omega^0)
\end{equation} 
at least for linearly-polarized fields. The more general case is not considered here.

\subsection{Subleading gravitational amplitudes}
\label{Sect:SubleadingGW}

Although there is no obstacle to considering scalar and electromagnetic wave propagation under wide classes of metric transformations, it is significantly more difficult to do so in the gravitational case. The reason for this is essentially that the $s=2$ transport and gauge conditions reviewed in Sect. \ref{Sect:Review} are derived \cite{HarteOptics1} using the vacuum Einstein equation (perhaps with a cosmological constant), so $g_{ab}$ and $\hat{g}_{ab}$ should both be vacuum solutions. This is not merely a technical restriction: If there is a nonzero stress-energy tensor associated with $g_{ab}$, it is necessarily perturbed by a passing gravitational wave. But those perturbations backreact, affecting the propagation of the wave. Details of this effect are not universal, but depend on precisely which type of matter is associated with the background stress-energy.

To be more precise about when such complications arise, let $\tau_{ab} = (8\pi)^{-1} (R_{ab} - \frac{1}{2} g_{ab} R)$ denote the  stress-energy tensor associated with $g_{ab}$. If a gravitational wave with metric perturbation $\Psi_{ab} = \Re \psi_{ab}$ is considered, the stress-energy tensor is perturbed to $\tau_{ab}+\delta \tau_{ab}$. Einstein's equation implies that $\tau_{ab}$ must be conserved with respect to $g_{ab}$ and also that $\tau_{ab} + \delta \tau_{ab}$ must be conserved with respect to $g_{ab} + \Psi_{ab}$. Schematically, it follows that
\begin{equation}
	\nabla^a \delta \tau_{ab} \sim ``\tau \nabla \Psi + \Psi \nabla \tau,"
\end{equation} 
so $\delta \tau_{ab}$ is generically nonzero when $\tau_{ab} \neq 0$. Noting that $\psi_{ab}$ admits the high-frequency expansion \eqref{WKB}, this also suggests\footnote{Stress-energy conservation constrains only the divergence of $\delta \tau_{ab}$. The remaining portions of this perturbation must be determined by the detailed equations of motion associated with the matter involved, and it is possible for divergence-free components to scale differently with $\omega$.} that $\delta \tau_{ab}(x;\omega) = \Re [\delta \tau^0_{ab}(x) e^{i \omega \varphi(x)}] + \mathcal{O}(\omega^{-1})$. The subleading transport equation for $\calA^1_{ab}$ would thus be modified from its vacuum counterpart, though not necessarily the leading-order equation associated with $\calA^0_{ab}$.

It follows that the geometric-optics results of Sect. \ref{Sect:GOgrav} are not necessarily affected by the presence of background matter. However, those aspects of Sect. \ref{Sect:inheritGW} which go beyond geometric optics are; they should be understood as restricted to transformations in which $g_{ab}$ and $\hat{g}_{ab}$ are both vacuum. The transformation laws for the Newman-Penrose scalars implicitly take the vacuum assumption into account, in that the background expressions for these quantities had already been simplified in Paper I using the vacuum Einstein equation. Nevertheless, the vacuum assumption is not explicitly used in any of the calculations above. By contrast, finding a transformation rule $\calA^1_{ab} \mapsto \hat{\calA}^1_{ab}$ for the subleading gravitational amplitudes appears to require that the vacuum Einstein equation be applied in an essential way. This adds significant complication and is left for later work.

\section{Generating new solutions}
\label{Sect:Examples}

One application of our metric-invariance results is that they allow new solutions to be generated from old ones---for scalar, electromagnetic, or gravitational waves. This is most straightforward when the old metric $g_{ab}$ and the new metric $\hat{g}_{ab}$ are related via \eqref{gTilde}, although it is possible to use diffeomorphism invariance to also make interesting statements in more general contexts. This section discusses some simple examples which extend a spherical-wave solution in flat spacetime to a variety of other geometries, without using diffeomorphism invariance. Sect. \ref{Sect:Scattering} below provides a more-complicated example which does use diffeomorphism invariance.

\subsection{Spherical waves in flat spacetime}
\label{Sect:sph}

Perhaps the simplest high-frequency solutions in flat spacetime are those with planar wavefronts. However, it is somewhat more interesting to consider spherical waves, and that is what we do here. Starting with ordinary spherical coordinates $(t,r,\Theta, \Phi)$ on a Minkowski background $g_{ab}$, it is convenient to introduce the radial null coordinates $u \equiv t-r$ and $v \equiv t+r$ so
\begin{equation}
	\rmd s^2 = g_{\mu\nu} \rmd x^\mu \rmd x^\nu = -\rmd u \rmd v + r^2 (\rmd\Theta^2 + \sin^2 \Theta \rmd\Phi^2).
	\label{dsFlat}
\end{equation} 
The $u$ coordinate here has the interpretation of a retarded time, so one valid eikonal for an outgoing spherical wave is
\begin{equation}
	\varphi = u .
	\label{eikonalFlat}
\end{equation}
The rays are then tangent to $k^\mu \partial_\mu = 2 \partial_v$. They are diverging but shear-free:
\begin{equation}
	\nabla \cdot k = \frac{2}{r}, \qquad \sigma = 0.
\end{equation}
The $n=0$ scalar transport equation \eqref{xPort} is easily solved in this case to yield the amplitudes
\begin{equation}
	\calA_0 = \frac{\alpha}{ r },
	\label{A0flat}
\end{equation}
where $\alpha=\alpha(u,\Theta,\Phi)$ is arbitrary and encodes a waveform for each ray. Furthermore, one solution for the subleading amplitude is
\begin{equation}
	\calA_1 = i ( \nabla^2_{\perp}  + \mu^2 r^2 ) \frac{\calA_0}{2r},
	\label{A1flat}
\end{equation}
where 
\begin{equation}
	\nabla^2_{\perp} \equiv \csc^2\Theta [\sin \Theta \partial_\Theta(  \sin \Theta \partial_\Theta) + \partial_\Phi^2] 
\end{equation}
denotes the Laplacian on a unit 2-sphere. The field mass $\mu$ thus induces a secularly-growing phase shift as $r$ increases. By contrast, the mass-independent portion of $\calA_1$ decays more rapidly at large distances than its leading-order counterpart.

Many observables are connected with $\langle T_{ab} \rangle$, and it follows from \eqref{TSc1} that this can be constructed, through leading and subleading orders, from $| \calA_0 + \omega^{-1} \calA_1 |^2$ and $k^\mathrm{cor}_a$. For the spherical waves just described,
\begin{equation}
	| \calA_0 + \omega^{-1} \calA_1 |^2 = \frac{ 1 }{ r^2 } \left[ |\alpha|^2  +\frac{ 1 }{ \omega r } \Im (\alpha \nabla^2_\perp \bar{\alpha}) + \mathcal{O}(\omega^{-2}) \right] 
\end{equation}
and
\begin{equation}
	k^\mathrm{cor}_a = k_a - \omega^{-1} \nabla_a \arg \alpha . 
\end{equation}
It follows that, e.g., the direction associated with the momentum density is frequency dependent and non-radial whenever there is a phase gradient across neighboring rays. 

These results are easily used to write down high-frequency approximations for electromagnetic and gravitational waves with spherical wavefronts, at least to leading order. Introducing the normalized basis element
\begin{equation}
	m_\mu \rmd x^\mu = \frac{ r }{ \sqrt{2}} ( \rmd \Theta + i \sin\Theta \rmd\Phi),
	\label{mDefFlat}
\end{equation}
it follows from \eqref{A0s}, \eqref{eExpand}, \eqref{eExpandGrav}, and \eqref{A0flat} that 
\begin{gather}
	\calA^0_a = \frac{ \alpha }{ r} ( e_+ m_a + e_- \bar{m}_a ) ,
	\qquad
	\calA^0_{ab} = \frac{ \alpha }{ r} ( e_+ m_a m_b + e_- \bar{m}_a \bar{m}_b) ,
	\label{A0highs}
\end{gather}
are valid $s=1$ and $s=2$ amplitudes if the polarization components $e_\pm$ are independent of the advanced time $v$. The products $\alpha e_+$ and $\alpha e_-$ are interpreted as waveforms for the two circularly-polarized basis components, and in terms of the coordinates, 
\begin{subequations}
\label{A0EMGravFlat}
\begin{gather}
	\calA^0_\mu \rmd x^\mu = \frac{ \alpha  }{ \sqrt{2} } [ (e_+ + e_-) \rmd \Theta + i (e_+ - e_-) \sin\Theta \rmd\Phi],
	\\
	\calA^0_{\mu\nu} \rmd x^\mu \rmd x^\nu = \alpha r  \left[ \tfrac{1}{2} (e_+ + e_-) (\rmd\Theta^2 - \sin^2 \Theta \rmd\Phi^2) + i (e_+ - e_-)  \sin \Theta \rmd\Theta d\Phi\right]. 
\end{gather}
\end{subequations}
Eqs. \eqref{fExpand} and  \eqref{RiemExpand} may also be used to write down the corresponding electromagnetic field and curvature perturbations:
\begin{subequations}
\begin{gather}
	f_{ab} = - 2 i \omega e^{i \omega \varphi} \calA_0 k_{[a} ( e_+ m_{b]} + e_- \bar{m}_{b]} ) + \mathcal{O}(\omega^0),
	\label{fFlat}
	\\
	\delta r_{abcd} = - 2 \omega^2 e^{i \omega \varphi} \calA_0 k_{[a} (e_+ m_{b]} m_{[c} + e_- \bar{m}_{b]} \bar{m}_{[c} ) k_{d]} + \mathcal{O}(\omega).
\label{curvFlat}
\end{gather}
\end{subequations}
For simplicity, we do not give examples of higher-order corrections when $s > 0$.

All of these results are associated with the flat line element \eqref{dsFlat}. If $g_{ab}$ is now transformed to $\hat{g}_{ab}$ via \eqref{gTilde}, the corresponding line elements are given by
\begin{align}
	\rmd \hat{s}^2 =  \Omega^2 \Big\{ V \rmd u^2 -\rmd u \left[ (1 + w_v) \rmd v + w_\Theta \rmd\Theta + w_\Phi \rmd\Phi\right] + \frac{ r^2 }{ 1 - \frac{1}{4} |Y|^2 } \Big[|1+\tfrac{1}{2} Y|^2 \rmd\Theta^2
	\nonumber
	\\
	~ +2 \Im Y \sin \Theta \rmd\Theta \rmd\Phi 
	 + |1-\tfrac{1}{2} Y|^2 \sin^2\Theta \rmd\Phi^2 \Big] \Big\},
	 \label{gTildeFlat}
\end{align}
where $V \equiv - w_u$ in order to be consistent with the notation in, e.g., \eqref{gConfKS}. It is clear that by suitably adjusting the deformation functions $\Omega$, $w_a$, and $Y$, all metric components except for $\hat{g}_{vv}$, $\hat{g}_{v\Theta}$, and $\hat{g}_{v\Phi}$ can be varied essentially at will. 

\subsection{Flat to flat transformations}
\label{Sect:Flatflat}

The background metric $g_{ab}$ here is flat and the transformed metrics $\hat{g}_{ab}$ are in general curved. There are however exceptions in which $\hat{g}_{ab}$ remains flat, and in these cases, transformations of the optical fields effectively generate new solutions from old ones in the same spacetime.

As a simple example, the angular coordinates $\Theta$ and $\Phi$ can be ``made Cartesian'' (at least in a finite region) using the deformation functions
\begin{equation}
\label{defCart}
\begin{gathered}
	\Omega^2 = \frac{ \csc \Theta }{ r^2 }  \qquad Y = - 2 [ \cos(\Theta/2) - \sin(\Theta/2)]^4 \sec^2 \Theta,
	\\
	w_\mu \rmd x^\mu = -( 1-r^2 \sin \Theta ) \rmd v.
\end{gathered}
\end{equation}
Substituting these choices into \eqref{gTildeFlat} results in the line element $\rmd \hat{s}^2 = - \rmd u \rmd v+\rmd \Theta+\rmd\Phi^2$, so $\Theta$ and $\Phi$ are now interpreted not as angles, but as Cartesian coordinates transverse to the wavefronts. Recalling \eqref{mapScalar}, this transformation maps a spherical wave with amplitude \eqref{A0flat} into a \textit{plane}-fronted wave with amplitude
\begin{equation}
	\hat{\calA}_0 =  \calA_0/\Omega =  (\sin \Theta)^{\frac{1}{2}} \alpha. 
	\label{AscalarPlane} 
\end{equation}
Moreover, the factor of $(\sin \Theta)^{ \frac{1}{2} }$ here may be absorbed into a redefinition of $\alpha(u,\Theta,\Phi)$; it is physically irrelevant.

Applying this transformation not only for scalar amplitudes, but also for their electromagnetic and gravitational counterparts requires $\hat{m}_a$. Given \eqref{defCart} and \eqref{theta}, it is possible to choose $\theta = 0$ in \eqref{mTilde}, implying that \eqref{mDefFlat} transforms to $\hat{m}_\mu \rmd x^\mu = 2^{- \frac{1}{2} } (\rmd\Theta + i \rmd\Phi)$. Applying \eqref{hate} and \eqref{eHatGrav} to \eqref{A0EMGravFlat} then results in what one might expect for plane-fronted electromagnetic and gravitational waves [other than the same redundant factor of $(\sin \Theta)^{\frac{1}{2}}$]:
\begin{equation}
\label{PW}
\begin{gathered}
	\hat{\calA}^0_\mu \rmd x^\mu =  \frac{ 1 }{ \sqrt{2}} (\sin \Theta)^{\frac{1}{2}}  \alpha[(e_+ + e_-) \rmd \Theta + i (e_+ - e_-) \rmd \Phi ],
	\\
	\hat{\calA}^0_{\mu\nu} \rmd x^\mu \rmd x^\nu = (\sin \Theta)^{\frac{1}{2}}  \alpha \left[ \tfrac{1  }{ 2 }  (e_+ + e_-) (\rmd \Theta^2 - \rmd \Phi^2) + i (e_+ - e_-) \rmd \Theta \rmd \Phi \right]. 
\end{gathered}
\end{equation}
These are plane waves (as opposed to plane-fronted waves) when the $\alpha e_\pm$ waveforms are independent of $\Theta$ or $\Phi$.

As another example, the originally-spherical wavefronts associated with solutions in Sect. \ref{Sect:sph} can be ``made cylindrical'' using the deformation functions
\begin{equation}
\begin{gathered}
	\Omega^2 = 	\frac{\csc \Theta}{r} , \qquad Y = 2 - \frac{ 4 }{ 1 + r \sin\Theta } ,
	\\
	w_\mu \rmd x^\mu = - (1 + r \sin\Theta ) \rmd v.
\end{gathered}
\label{xformCyl}
\end{equation}
Substituting these expressions into \eqref{gTildeFlat} results in the line element $\rmd\hat{s}^2 = - \rmd u \rmd v + r^2 \rmd\Theta^2 + \rmd\Phi^2$, so $r= \frac{1}{2}(v-u)$ is now interpreted as a cylindrical radius,  $\Theta$ an angular coordinate around the cylinder, and $\Phi$ an elevation along the axis of that cylinder. The rays remain tangent to $\partial_v$ and therefore emanate radially from the axis of the cylinder. Moreover, \eqref{A0scalarXform} implies that
\begin{equation}
	\hat{\calA}_0 = \left(\frac{ \sin \Theta }{ r } \right)^{\frac{1}{2}} \alpha.
\end{equation}
As expected, this falls off with the square root of the distance away from the axis of the cylinder. The higher-spin amplitudes are easily written down by using \eqref{mTilde} to see that now, $\hat{m}_\mu \rmd x^\mu = 2^{- \frac{1}{2} }  (r \rmd\Theta + i \rmd\Phi)$.

Note that none of these results require the solution to any differential equations; even quite drastic changes in the geometry can be understood using purely-algebraic transformations. Of course, it would have been straightforward to instead derive amplitudes for planar and cylindrical waves directly from the transport equations. Much more complicated flat-to-flat deformations are however possible, and direct calculation may be less clear in those cases. Further investigation is nevertheless required. Although conformal transformations which map flat metrics into flat metrics are understood---comprising dilatations, inversions, and Poincar\'{e} transformations \cite{Kastrup2008}---much less is known regarding the freedom to vary $w_a$ and $Y$. See however \cite{SenovillaKS, SenovillaBiconformal} for related results.

\subsection{Cosmological spacetimes}
\label{Sect:ExConform}

The simplest transformations to non-flat spacetimes may be generated using only the conformal degree of freedom in \eqref{gTilde}, and the most interesting examples in this class are cosmological. All Friedmann-Lema\^{i}tre-Robertson-Walker (FLRW) metrics are conformally flat, meaning that at least in finite regions, there exist coordinates $(u,v,\Theta,\Phi)$ such that
\begin{equation}
	\rmd \hat{s}^2 = \Omega^2 \rmd s^2= \Omega^2 [ -\rmd u \rmd v + r^2 (\rmd\Theta^2 + \sin^2 \Theta \rmd\Phi^2)]. 
\end{equation}
Explicit decompositions of this form are known for general FLRW metrics \cite{confFRW, confFRW2}, although it is common to use them only when the homogeneous constant-time hypersurfaces are spatially flat; $\Omega$ is then identified with the scale factor and depends only on $t = \frac{1}{2} (v+u)$. One example in this spatially-flat class is the de Sitter geometry, which may be described by
\begin{equation}
	\Omega = \frac{1}{ \sqrt{\Lambda/3} t },
	\label{dS}
\end{equation}
where $\Lambda > 0$ denotes the cosmological constant. Anti-de Sitter metrics with $\Lambda < 0$ may be generated by instead letting, e.g., $\Omega =  (\sqrt{-\Lambda/3} r \cos \Theta)^{-1}$. 

Regardless of precisely which conformal factor is used, it follows from \eqref{A0scalarXform} and \eqref{A1Hat} that the $s=0$ flat-spacetime, spherical-wave amplitudes \eqref{A0flat} and \eqref{A1flat} are replaced by
\begin{equation}
	\hat{\calA}_0 = \Omega^{-1} \left( \frac{  \alpha }{ r } \right), \qquad \hat{\calA}_1 =   i \Omega^{-1}  ( \nabla^2_\perp + \mu^2 r^2 -2r \vartheta ) \frac{ \alpha }{ 2 r^2}
	\label{A1HatEx}
\end{equation}
in FLRW spacetimes, where $\vartheta$ satisfies \eqref{varthetaEvolve}. In the spatially-flat case, 
\begin{align}
	\partial_t \vartheta = \tfrac{1}{2} [ (1-6 \xi) \partial_t^2 \Omega/\Omega - \mu^2 (\Omega^2-1)].
\end{align}
and for a de Sitter metric in which $\Omega$ is given by \eqref{dS}, one solution is
\begin{equation}
	\vartheta = \tfrac{1}{2} \mu^2 \left( t + \frac{3 }{ \Lambda t } \right)  -\frac{ (1-6 \xi) }{ t }.
\end{equation}
More generally, it follows from \eqref{Tsc2} that $\langle \hat{T}_{ab} \rangle = \Omega^{-2} \langle T_{ab} \rangle + \mathcal{O}(\omega^0)$ for scalar fields in any FLRW metric. 

Electromagnetic and gravitational waves may be understood similarly. It is immediate from \eqref{f0HatSimp} and \eqref{RiemXform} that the electromagnetic field \eqref{fFlat} is preserved as-is while the curvature perturbation \eqref{curvFlat} associated with a gravitational wave changes only by an overall factor of $\Omega$. Our results make the propagation of high-frequency gravitational waves in cosmological contexts essentially trivial.

\subsection{Spherically-symmetric metrics}
\label{Sect:exampleSph}

It is possible to obtain any static (and at least some non-static) spherically-symmetric geometry by applying Kerr-Schild and conformal transformations to a flat metric \cite{confKS}; using such a metric together with the outgoing $k_a$ associated with the eikonal \eqref{eikonalFlat}, spherically-symmetric line elements can be put into the form 
\begin{equation}
	\rmd \hat{s}^2 = \Omega^2 [V \rmd u^2 - \rmd u \rmd v + r^2 (\rmd\Theta^2 + \sin^2 \Theta \rmd\Phi^2) ].
	\label{dsconfKS}
\end{equation}
For example, the Schwarzschild solution with mass $M$ results from $\Omega=1$ and $V=2M/r$. De Sitter and anti-de Sitter metrics result from\footnote{This gives a different representation for the de Sitter and anti-de Sitter metrics than the purely-conformal one mentioned in Sect. \ref{Sect:ExConform}.} $\Omega =1$ and $V = (\Lambda/3)r^2$. The Schwarzschild-de Sitter geometry arises by simply adding together these two forms for $V$, providing an example of the aforementioned linearity of Einstein's equation for Kerr-Schild metric perturbations. Nontrivial forms for $\Omega$ arise in, e.g., the metrics associated with stellar interiors.

Regardless of the precise forms for $\Omega$ and $V$, it follows from the results of Sects. \ref{Sect:Rays} and \ref{Sect:Geo} that the rays are preserved, the scalar amplitude \eqref{A0flat} is replaced by $\calA_0/\Omega$, the leading-order electromagnetic field \eqref{fFlat} remains as-is, and the leading-order curvature perturbation \eqref{curvFlat} is multiplied by $\Omega$. Going one order beyond geometric optics, it follows from \eqref{A1Hat} that the scalar $\calA_1$ which is given by \eqref{A1flat} is replaced by the $\hat{\calA}_1$ in \eqref{A1HatEx}. Furthermore, the electromagnetic $\calA^1_a$ is related to its hatted counterpart by \eqref{A1HatEM}. Noting that $\sigma = 0$, it follows from \eqref{varthetaEM} that the electromagnetic phase shift governed by $\vartheta_{a}{}^{b}$ can be chosen to vanish for any spherical wave in any spherically-symmetric metric with the form \eqref{dsconfKS}. The scalar case is more complicated, as the phase shift governed by $\vartheta$ is in general nontrivial for spherically-symmetric metrics. Using \eqref{varthetaEvolve} and the Kerr-Schild representation for the Schwarzschild-de Sitter metric given in the previous paragraph, one solution in that case is
\begin{equation}
	\vartheta =  (1-2 \xi)(\Lambda/3) r + (1-8 \xi) \frac{  M }{ 2 r^2 }.
\end{equation}

The physical interpretation of this discussion is that except for simple rescalings by conformal factors, the geometric optics associated with radially-outgoing---but not necessarily spherically symmetric---scalar, electromagnetic, and gravitational waves is unaffected by (at least static) spherically-symmetric deformations of the metric. Of course, this is valid in the particular representation given here for spherically-symmetric metrics. Similar invariance results do not hold in other, more-common representations, although appropriate transformations are easily derived.

As an example of such a transformation, again consider the Schwarzschild case (but now without a de Sitter component). While the $r = \tfrac{1}{2}(v-u)$ appearing in \eqref{dsconfKS} corresponds to the ordinary Schwarzschild radius, $t = \tfrac{1}{2} (v+u)$ is not the time coordinate $t_\mathrm{Sch}$ which is most commonly associated with this geometry; the standard line element 
\begin{equation}
	\rmd \hat{s}^2 = - (1-2M/r) \rmd t_\mathrm{Sch}^2 + (1-2M/r)^{-1} \rmd r^2 + r^2 (\rmd \Theta^2 + \sin^2 \Theta \rmd \Phi^2)
	\label{gSchw}
\end{equation}
instead arises by defining
\begin{equation}
	t_\mathrm{Sch} \equiv t + 2 M \ln (r-2M).
\end{equation}
In terms of this, the eikonal \eqref{eikonalFlat} expands to
\begin{equation}
	\varphi = t_\mathrm{Sch} - r -  2 M \ln (r-2M)
\end{equation}
and the waveform $\alpha(u,\Theta, \Phi)$ which appears in the various amplitudes simply has $u$ replaced by this same combination of $t_\mathrm{Sch}$ and $r$. Although one might be tempted to associate with \eqref{gSchw} the flat line element $-\rmd t_\mathrm{Sch}^2 + \rmd r^2 + r^2 ( \rmd \Theta^2 + \sin^2 \Theta \rmd \Phi^2)$, this is not the ``correct'' identification; optical solutions are not preserved in the expected way. One must instead use
\begin{align}
	ds^2 &= - \rmd t^2 + \rmd r^2 + r^2 (\rmd \Theta^2 + \sin^2 \Theta \rmd \Phi^2 )
	\nonumber
	\\
	& = -\left[ \rmd t_\mathrm{Schw} - \left( \frac{ 2 M}{ r - 2M } \right) \rmd r  \right]^2 + \rmd r^2 + r^2 (\rmd \Theta^2 + \sin^2 \Theta \rmd \Phi^2 )
\end{align}
which is the flat line element associated with the Kerr-Schild transformation \eqref{dsconfKS}. This is the ``obvious'' identification in the coordinates $(t,r,\Theta,\Phi)$, but not in the coordinates $(t_\mathrm{Schw}, r, \Theta, \Phi)$.

\subsection{Robinson-Trautman metrics}

The class of transformed line elements \eqref{gTildeFlat} is very broad, and in many cases, it straightforward to look through collections of exact solutions to Einstein's equations (e.g., \cite{ExactSolns, GriffithsExact}) and to match deformation functions essentially by eye. 

As an example, this is possible for the Robinson-Trautman metrics, which may be characterized mathematically as those geometries which admit a null geodesic congruence which is twist-free and shear-free but with nonzero expansion. The Robinson-Trautman line elements may be written as \cite{GriffithsExact}
\begin{equation}
	\rmd \hat{s}^2 = - \rmd u \rmd v - 2 H \rmd u^2+ \frac{ v^2 }{ 2 P^2 } \rmd\zeta \rmd\bar{\zeta},
\end{equation}
where $\zeta$ is a complex stereographic coordinate, $P = P(u, \zeta, \bar{\zeta})$, and $H = H(u,v,\zeta,\bar{\zeta})$ depends on $P$ and other quantities in a known way. 

To see how the Robinson-Trautman metrics may be generated from our flat line element \eqref{dsFlat}, first relate $\zeta$ to $\Theta$ and $\Phi$ via $\zeta = e^{i \Phi} \cot (\Theta/2)$, so
\begin{equation}
	\rmd \zeta \rmd \bar{\zeta} = \tfrac{1}{4} \csc^4 (\Theta/2) ( \rmd \Theta^2 + \sin^2 \Theta \rmd \Phi^2).
	\label{stereo}
\end{equation}
It then follows from comparison with \eqref{gTildeFlat} that the appropriate deformation functions are
\begin{equation}
\begin{gathered}
	\Omega = \frac{ v}{ \sqrt{8} P r \sin^2 (\Theta/2) } 
	, \qquad 
	Y= 0,
	\\
	w_\mu \rmd x^\mu = \frac{ 8 P^2 r^2  }{ v^2 }(2 H \rmd u + \rmd v) \sin^4 (\Theta/2)  - \rmd v.
\end{gathered}
\end{equation}
At least in geometric optics, it is thus straightforward to transform amplitudes and field strengths associated with spherical waves in flat spacetime into amplitudes and field strengths in arbitrary Robinson-Trautman spacetimes. The optical rays then coincide with the shear-free, twist-free congruence picked out by the Robinson-Trautman class.

\subsection{Gravitational wave backgrounds}

It is also straightforward to transform from a flat metric to a Kundt metric, which is defined to be a geometry in which there exists a null geodesic congruence free of twist, expansion, and shear. The Kundt line elements may be written in the form \cite{GriffithsExact}
\begin{equation}
	\rmd \hat{s}^2 = - \rmd u ( \rmd v + 2 H \rmd u + 2 W \rmd \zeta + 2 \bar{W} \rmd \bar{\zeta} ) + 2 P^{-2} \rmd \zeta \rmd \bar{\zeta},
	\label{Kundt}
\end{equation}
where $\zeta$ is again a complex stereographic coordinate, $P = P(u, \zeta, \bar{\zeta})$ and $H = H(u, v, \zeta , \bar{\zeta})$ are real, and $W = W(u,v,\zeta,\bar{\zeta})$ may be complex. Special cases include gravitational plane waves---and more generally \textit{pp}-waves---as well as, e.g., solutions which may be interpreted as gravitational waves propagating on de Sitter or anti-de Sitter backgrounds. 

Using \eqref{stereo} and comparing \eqref{gTildeFlat} with \eqref{Kundt} shows that the Kundt line elements may be generated from the flat line element \eqref{dsFlat} using the deformation functions
\begin{equation}
	\Omega = \frac{ 1 }{ \sqrt{2} P r \sin^2 (\Theta/2) } 
	, \qquad 
	Y= 0,
\end{equation}
and
\begin{align}
	w_\mu \rmd x^\mu = 2 P^2 r^2  [ 2 H \rmd u + \rmd v - 2 \csc^2 (\Theta/2) ( \Re W \cos \Phi - \Im W \sin \Phi) \rmd \Theta
	\nonumber
	\\
	~ +4 \cot (\Theta/2) (\Re W \sin \Phi - \Im W \cos \Phi ) \rmd \Phi]\sin^4 (\Theta/2)  - \rmd v.
\end{align}
Again, the transformation results above may be used to carry over flat amplitudes at least in geometric optics to amplitudes in arbitrary Kundt spacetimes. 

It can be somewhat more natural to start with plane-fronted fields in flat spacetime instead of spherical ones: Letting $\zeta = \Theta + i \Phi$ instead of $e^{i \Phi} \cot(\Theta/2)$, it follows that in this case, \eqref{stereo} is replaced by $\rmd \zeta \rmd \bar{\zeta} = \rmd \Theta^2 + \rmd \Phi^2$ and
\begin{equation}
\begin{gathered}
	\Omega = \frac{\sqrt{2}}{P} , \qquad \quad Y=0, 
	\\
	w_\mu \rmd x^\mu = \tfrac{1}{2} P^2 ( 2 H \rmd u + \rmd v + \Re W \rmd \Theta - \Im W \rmd \Phi) - \rmd v.
\end{gathered}
\end{equation}
That these functions exist physically implies that plane-fronted optical fields in flat spacetime are unaffected by Kundt waves which ``propagate in the same direction,'' in the sense that the preferred null congruence $\partial_v$ associated with the deformed metric is tangent to the optical rays. This implies in particular that plane-fronted gravitational waves, which may be characterized by $P=2^{-\frac{1}{2}}$ and $W=0$, do not affect the high-frequency propagation of scalar, electromagnetic, or (additional) gravitational waves which propagate in the same direction. In the electromagnetic case, this is related to an exact result obtained in \cite{HarteEMKS}. Qualitatively, it is also similar to the statement in Sect. \ref{Sect:exampleSph} that spherical waves are unaffected by spherically-symmetric metrics. 

As in the spherical case, this insensitivity of plane-fronted optical fields to gravitational waves must be understood in the context of the line elements \eqref{dsFlat} and \eqref{Kundt}. It is not apparent in, e.g., the transverse-traceless gauge which is more commonly used to (approximately) describe gravitational radiation. This may be understood physically as due to the fact that the transverse-traceless gauge is chosen to hold fixed a particular family of timelike geodesics, but timelike geodesics do not have the same metric-invariance properties as high-frequency fields. Nevertheless, transformations to transverse-traceless gauge are known \cite{HarteVines} and may be used together with our invariance result to  find transformed amplitudes in that gauge.

\section{A scattering problem}
\label{Sect:Scattering}

Although the metric transformations \eqref{gTilde} allow a wide variety of geometries to be generated from a given background, there are limitations. Indeed, if a metric is perturbed using the common gauges associated with perturbation theory in general relativity, it is only in special cases that the perturbed and background metrics will be related by such a transformation. Nevertheless, as stated at the end of Sect. \ref{Sect:Rays}, diffeomorphisms can be used to bring any given metric into the appropriate form---at least in finite regions. These diffeomorphisms are relatively simple to find in a perturbative context. There, they would more often be referred to as gauge transformations associated with the freedom to identify points in different ways in the perturbed and background spacetimes.

We now consider an explicit example in which the gravitational deflection of light by a point mass is computed not by solving geodesic or other transport equations, but by finding a gauge transformation which allows a plane wave in flat spacetime to be deformed into a scattered wave in a nontrivial geometry. Only the leading-order geometric-optics result is considered, and terms are retained only to leading order in the mass of the central object. 

\subsection{Preliminaries}

Beginning in a general context with a background metric $g_{ab}$, suppose that we are interested in fields propagating on the perturbed family of spacetimes with metrics
\begin{equation}
	\tilde{g}_{ab} = g_{ab} + \epsilon h_{ab},
	\label{gPert}
\end{equation}
where $\epsilon \ll 1$ is a bookkeeping parameter. Also introduce a 1st-order gauge vector $\xi^a$ which generates an $\epsilon$-dependent family of diffeomorphisms $\phi$ via
\begin{equation}
	(\phi_* \tilde{g})_{ab} = \tilde{g}_{ab} + \epsilon \mathcal{L}_\xi g_{ab}	.
	\label{phigTilde}
\end{equation}
Given $g_{ab}$ and an associated eikonal $\varphi$, we would like to construct $\xi^a$ such that the rays are the same (and null) whether they're computed using $g_{ab}$ or $(\phi_* \tilde{g})_{ab}$. This amounts to enforcing \eqref{diffeo1}. Contracting that equation first with $k^b$ results in the transport equation
\begin{equation}
	\mathcal{L}_k ( k \cdot \xi) = - \tfrac{1}{2} k^a k^b h_{ab}
	\label{kxi}
\end{equation}
for $k \cdot \xi$. Contracting it with $m^b$ while assuming that this is parallel transported with respect to $g_{ab}$ results in
\begin{equation}
	\mathcal{L}_k ( m \cdot \xi) = -k^a m^b h_{ab} - m \cdot \nabla (k \cdot \xi) + m^a \xi^b \nabla_a k_b ,
	\label{mxi}
\end{equation}
which is a transport equation for $m \cdot \xi$. Contraction with $n^b$ yields no additional information.

\subsection{Transforming the metric}

Now specialize so that $g_{ab}$ is a flat metric and the linear perturbation $h_{ab}$ may be interpreted as that due to a pointlike object with mass $\epsilon M$ in Lorenz gauge. Using inertial Minkowski coordinates $(t,x,y,z)$ on $g_{ab}$ with respect to which this mass is static, the ``Newtonian'' metric perturbation is \cite{Wald}
\begin{equation}
	h_{ab} = \frac{2 M}{r} (g_{ab} + 2 \nabla_a t \nabla_b t),
	\label{hNewt}
\end{equation}
where $r = \sqrt{x^2+y^2+z^2}$ is the usual radial coordinate and the $z$ coordinate used here is not to be confused with the $z$ in \eqref{zDef}. 

Plane-fronted optical fields on flat backgrounds have already been discussed in Sect. \ref{Sect:Flatflat}, although the notation there is unconventional. Changing $r$ to $x$, $\Theta$ to $y$, and $\Phi$ to $z$, optical fields in the $s=0$ case can be described by
\begin{align}
	\varphi = t- x, \qquad \calA_0 = \calA_0 ( t-x,y,z).
	\label{eikCart}
\end{align}
The higher-spin cases follow trivially from this together with $m_\mu \rmd x^\mu = 2^{-\frac{1}{2}} (\rmd y + i \rmd z)$. While these fields are valid solutions to the equations of geometric optics associated with $g_{ab}$, they are not solutions associated with $\tilde{g}_{ab}$.

Our transformation rules may nevertheless be applied by constructing diffeomorphisms $\phi$ in which $(\phi_*\tilde{g})_{ab}$ satisfies \eqref{diffeo1}. An associated gauge vector satisfies \eqref{kxi} and \eqref{mxi}, and introducing an integration constant $r_0 > 0$, one class of solutions is given by
\begin{equation}
	k \cdot \xi = 2M \ln \left( \frac{ r-x}{r_0} \right) , \qquad m \cdot \xi = - \sqrt{2} M \left(\frac{  r+x}{ y-i z}\right).
\end{equation}
More explicitly, 
\begin{align}
	\xi^\mu \partial_\mu = -M \ln \left( \frac{ r-x}{r_0} \right) (\partial_t - \partial_x) - \frac{ 2 M }{ r-x } ( y \partial_y + z \partial_z).
	\label{xiEx}
\end{align}
Different choices for $r_0$ correspond to gauge vectors which differ only by multiples of a background Killing vector, and are not particularly interesting. Other gauge vectors which differ in more complicated ways are also possible, and these can produce transformed fields with different physical interpretations. The choice adopted here results in a transformed field whose rays are initially planar and traveling in the $+x$ direction; the transverse components of $\xi^\mu$ vanish as $x \to -\infty$.

If $(\phi_* \tilde{g})_{ab}$ is now computed using \eqref{gPert}, \eqref{phigTilde}, \eqref{hNewt}, and \eqref{xiEx}, it takes the form \eqref{gTilde} and may be identified with $\hat{g}_{ab}$ for some deformation functions $\Omega$, $w_a$, and $Y$. Although complete knowledge of these functions is not needed to transform the optical fields, we list them here for completeness:
\begin{equation}
\label{exDef}
\begin{gathered}
	w_\mu dx^\mu = - \frac{ 2\epsilon M}{ r} \left[ dt + \frac{ ydy+zdz }{ r-x } \right] + \mathcal{O}(\epsilon^2) , 
	\\
	 \Omega = 1 + \mathcal{O}(\epsilon^2), \qquad 
Y = \frac{2\epsilon M}{r} \left( \frac{  r+x }{ y-i z} \right)^2 + \mathcal{O}(\epsilon^2).
\end{gathered}
\end{equation}
These expressions become trivial as $x \to -\infty$ but diverge as $x \to + \infty$, implying that as claimed, the incoming waves are asymptotically planar while the outgoing ones are not. 

\subsection{Fields associated with $(\phi_* \tilde{g})_{ab}$}

By construction, the eikonal is preserved as-is by the transformation $g_{ab} \mapsto (\phi_* \tilde{g})_{ab}$. Recalling \eqref{A0scalarXform} and the trivial conformal factor in \eqref{exDef}, scalar amplitudes are preserved as well. Results for electromagnetic or gravitational waves are somewhat more complicated, as the $m_a$ associated with $g_{ab}$ is not parallel propagated or correctly normalized with respect to $\hat{g}_{ab}=(\phi_* \tilde{g})_{ab}$. A replacement $\hat{m}_a$ is nevertheless given by \eqref{mTilde}. Noting that $\sigma = 0$ and $\mathcal{L}_k \arg Y = 0$, it follows from \eqref{theta} that it is possible to choose $\theta = 0$. Hence,
\begin{equation}
	\hat{m}_a = m_a + \tfrac{1}{2} Y \bar{m}_a + l k_a+  \mathcal{O}(\epsilon^2).
	\label{mPert}
\end{equation}
The undetermined scalar $l$ is of order $\epsilon$ and is to be fixed by demanding that $\hat{m}_a$ be parallel transported along the rays with respect to $(\phi_* \tilde{g})_{ab}$. There is no need to compute it, however.  From \eqref{f0Hat} and \eqref{detPert}, the leading-order electromagnetic field associated with $(\phi_* \tilde{g})_{ab}$ is independent of $l$ and governed by
\begin{align}
	\hat{\mathcal{F}}^0_{ab} = \calA_0 k_{[a} \left[ ( e_+ + \tfrac{1}{2} \bar{Y} e_- ) m_{b]}  
	 +(e_- +  \tfrac{1}{2} Y e_+  ) \bar{m}_{b]} \right] + \mathcal{O}(\epsilon^2).
	 \label{FhatEx}
\end{align}
Similarly, the leading-order curvature perturbation is governed by
\begin{align}
	\hat{\mathcal{R}}^0_{abcd} = \calA_0 k_{[a} \big[ e_+ m_{b]} m_{[c} + e_- \bar{m}_{b]} \bar{m}_{[c} + \tfrac{1}{2} ( e_+ Y + e_- \bar{Y} )
	 \nonumber
	 \\
	 ~ \times (m_{b]} \bar{m}_{[c} + \bar{m}_{b]} m_{[c} ) \big]  k_{d]} + \mathcal{O}(\epsilon^2).
	 \label{RhatEx}
\end{align}

\subsection{Fields in the original gauge}

We have now used plane-fronted waves on a Minkowski background to derive optical fields scattered by a point mass. However, the associated metric components are complicated and difficult to interpret. In practice, it can be more useful to apply $\phi^{-1}$ to $\hat{O}_0 = \{ (\phi_* \tilde{g})_{ab}; \varphi, \hat{\calA}^0_B\}$, resulting in optical fields $\tilde{O}_0 = \{ \tilde{g}_{ab}; \tilde{\varphi}, \tilde{\calA}^0_B \}$ which are associated with the original, simpler form for the perturbed metric. Applying $\phi^{-1}$ first to the eikonal \eqref{eikCart}, it follows from \eqref{xiEx} that
\begin{align}
	\tilde{\varphi} &= \varphi - \epsilon \mathcal{L}_\xi \varphi
	= t-x +2 \epsilon M \ln \left( \frac{ r-x }{ r_0 } \right) + \mathcal{O}(\epsilon^2).
\end{align}
The perturbation $2\epsilon M \omega \ln [(r-x)/r_0]$ acts as a kind of phase shift with respect to the flat-spacetime plane wave we started with. It diverges logarithmically if $x>0$ and $y,z \to 0$, which is where rays cross due to gravitational focusing. The high-frequency ansatz \eqref{WKB} breaks down there, as does the approximation \eqref{hNewt} for the point-mass metric and the expansion in powers of $\epsilon$. While the analysis can be modified to be more realistic and more accurate, we nevertheless proceed without introducing any additional complications.

\begin{figure}
	\centering
	\includegraphics[width=.5\linewidth]{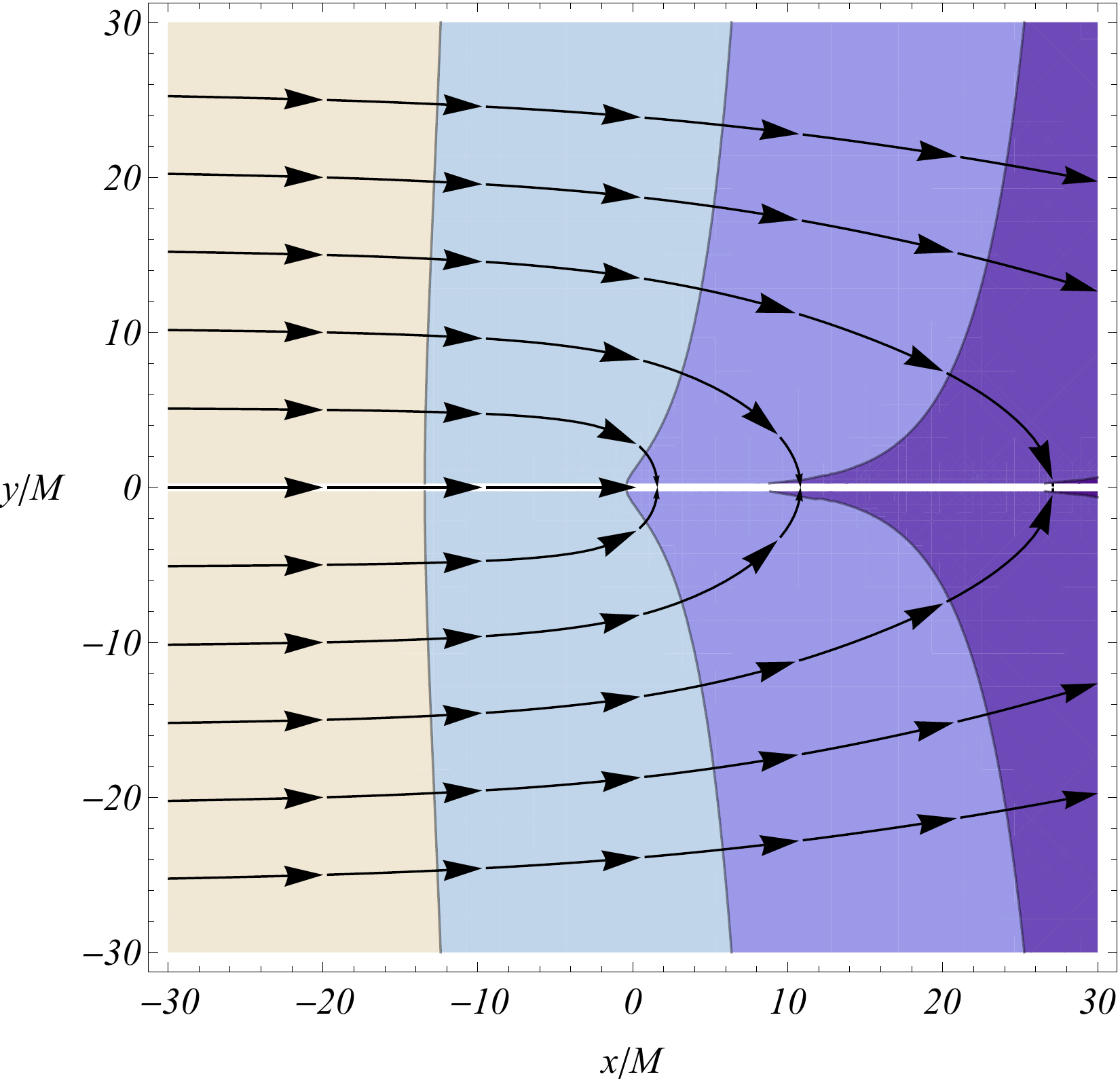}
	\caption{A $t=z=0$ contour plot of the scattered eikonal $\tilde{\varphi}$ together with the optical rays tangent to $\tilde{g}^{ab} \tilde{k}_b$. The central mass is at the $x=y=0$ origin and $\epsilon = 1$ and $r_0 = M$.}
	\label{fig}
\end{figure}

The next step is to determine how the rays are bent by the central mass. Given $\tilde{\varphi}$, a wavevector $\tilde{k}_a$ can be computed using $- \nabla_a \tilde{\varphi}$ or $k_a - \epsilon \mathcal{L}_\xi k_a$; the result is the same. Performing this computation and then raising the index shows that rays must be tangent to 
\begin{equation}
	(\tilde{g}^{\mu\nu} \tilde{k}_\nu) \partial_\mu = \left( 1+\frac{ 2 \epsilon M }{r} \right) \partial_t + \partial_x - \frac{ 2\epsilon M }{ r} \left( \frac{y \partial_y + z \partial_z}{r-x} \right) + \mathcal{O}(\epsilon^2).
\end{equation}
This and $\tilde{\varphi}$ are plotted in Fig. \ref{fig}. Some additional intuition for it may be gained by noting that as $x \to - \infty$, the rays become tangent to the background plane-wave vector field $\partial_t + \partial_x$. However, the rays become tangent to 
\begin{equation}
	\partial_t + \partial_x - 4 M \left( \frac{y \partial_y + z \partial_z}{y^2+z^2} \right)
\end{equation}
as $x \to + \infty$. The metric is trivial in these regions, so the change in angle of a light ray due to scattering, as seen by observers who are stationary with respect to the central mass, is $4 M /\sqrt{y^2+z^2}$. This matches the classical calculation for light bending by a point mass \cite{Wald}, although here it is obtained using a different method. It may also be noted that while the divergence of the congruence remains zero here, it follows from \eqref{sigDef} that the central mass induces a nonzero shear.

Amplitudes and field strengths may be found by similarly applying gauge transformations to the appropriate expressions in $\hat{O}_0$. For example, scalar amplitudes transform via
\begin{align}
	\tilde{\calA}_0 &=  \calA_0 - \epsilon \mathcal{L}_\xi \calA_0  = \left\{ 1 +  2 \epsilon M \left[ \ln \left( \frac{r-x}{r_0} \right)\partial_{t-x} +  \frac{ y \partial_y + z \partial_z }{r-x}  \right] \right\} \calA_0 + \mathcal{O}(\epsilon^2).
\end{align}
Although it is formally equivalent to the order in which we work, better behavior can sometimes be obtained by instead using $\xi^a$ to directly transform the arguments of of the original amplitude:
\begin{equation}
	\tilde{\calA}_0 (t-x,y,z) = \calA_0 \left( t-x +2 \epsilon M \ln \left( \frac{ r-x }{ r_0 } \right), y + \frac{ 2 \epsilon M y }{ r- x} ,  z + \frac{ 2 \epsilon M z }{ r- x} \right) + \mathcal{O}(\epsilon^2).
\end{equation}
This avoids additional errors incurred by expanding $\calA_0$.

Amplitudes for higher-spin fields differ in that they involve polarization tensors, and these too are affected by the central mass. The leading-order electromagnetic field strength can be computed using $\tilde{\mathcal{F}}^0_{ab} = \hat{\mathcal{F}}^0_{ab} - \epsilon \mathcal{L}_\xi \hat{\mathcal{F}}^0_{ab} + \mathcal{O}(\epsilon^2)$, where $\hat{\mathcal{F}}^0_{ab}$ is given by \eqref{FhatEx}. Similarly, the leading-order curvature perturbation due to a gravitational wave scattered by a point mass can be computed using $\tilde{\mathcal{R}}^0_{abcd} = \hat{\mathcal{R}}^0_{abcd} - \epsilon \mathcal{L}_\xi \hat{\mathcal{R}}^0_{abcd} + \mathcal{O}(\epsilon^2)$, where $\hat{\mathcal{R}}^0_{abcd}$ is given by \eqref{RhatEx}. Instead of displaying these results explicitly, more can be learned by instead computing how $\phi^{-1}$ acts on the transformed basis element $\hat{m}_a$: Using \eqref{mPert},
\begin{align}
	\tilde{m}_a = ( 1 + \epsilon M/r) m_a +  \sqrt{2} \epsilon M  \left( \frac{1+x/r }{ y-iz} \right) \nabla_a x + l k_a + \mathcal{O}(\epsilon^2).
\end{align}
If the irrelevant term containing $l$ is ignored, this reduces to $m_a$ as $x \to -\infty$ and 
\begin{equation}
	m_a + \left( \frac{ 2 \sqrt{2} \epsilon M  }{ y-iz } \right) \nabla_a x
\end{equation}
as $x \to +\infty$. That these limits differ implies that distant observers which are stationary with respect to the central mass would see an overall rotation in the polarization state of (say) a linearly-polarized electromagnetic field. However, this can largely be understood as due to the fact that the rays have been bent and polarization tensors must remain orthogonal to them.

The calculations in this section can be extended in a number of ways. One possibility would be to resolve the issues associated with ray crossings on the $x>0$, $y=z=0$ line. Another direction would involve repeating these calculations for more complicated mass distributions. This is very simple in the Newtonian spacetimes considered here, as everything is linear and translation invariant: The gauge vector \eqref{xiEx} is easily modified to act essentially as a Green function, and that can be convolved with arbitrary mass distributions. All of the field transformations then follow by straightforward differentiation of the resulting gauge vector.

\section{Discussion}

We have shown that individual solutions in geometric optics are compatible with a very large class of metrics. Given one metric $g_{ab}$ together with an associated eikonal $\varphi$, this eikonal and its rays are preserved by all transformations $g_{ab} \mapsto \hat{g}_{ab}$ in which $\hat{g}_{ab}$ has the form \eqref{gTilde}. Such transformations involve essentially arbitrary combinations of conformal transformations, determined by $\Omega$, extended Kerr-Schild transformations along the rays, determined by $w_a$, and pairs of complex Kerr-Schild transformations transverse to the rays, determined by $Y$. They preserve not only the rays, but also---up to conformal rescalings---the scalar amplitudes of geometric optics: $\calA_0 \mapsto \calA_0/\Omega$. Electromagnetic and gravitational amplitudes are modified in somewhat more complicated ways by these transformations, although observables which are insensitive to details of their polarization states are not. To leading order, averaged stress-energy tensors associated with scalar and electromagnetic waves are preserved up to conformal rescaling: $\langle T_{ab} \rangle \mapsto \Omega^{-2} \langle T_{ab} \rangle$. For gravitational waves, the average of the perturbed Bel-Robinson tensor is similarly preserved: $\langle \delta T_{abcd} \rangle \mapsto \Omega^{-2} \langle \delta T_{abcd} \rangle$. 

If the class of allowed metric transformations is restricted so that $Y=0$, even the details of a field's polarization state are preserved: To leading order, gradients of  scalar fields $\Psi = \Re \psi$, electromagnetic field strengths $F_{ab} = \Re f_{ab}$, and curvature perturbations $\delta R_{abcd} = \Re \delta r_{abcd}$ transform as
\begin{equation}
	\nabla_a \Psi \mapsto \Omega^{-1} \nabla_a \Psi, \qquad F_{ab} \mapsto F_{ab}, \qquad \delta R_{abcd} \mapsto \Omega \delta R_{abcd} 
	\label{fieldSummary}
\end{equation}
when the metric is deformed using arbitrary choices of $\Omega$ and $w_a$. Although the conformal factor appears explicitly here in the spin-0 and spin-2 cases, these results hold for a class of transformations which involve five free functions; the four functions associated with $w_a$ have no effect. Moreover, the applicability of these results (and those for which $Y \neq 0$) can be broadened even further by taking advantage of diffeomorphism invariance.

Metric invariance results which go beyond geometric optics depend on precisely which observable is considered. For example, the corrected scalar wavevector $k_a^\mathrm{cor}$ remains invariant under metric transformations generated by arbitrary $\Omega$, $w_a$, and $Y$. Nevertheless, the rays associated with this wavevector can differ when computed using $g_{ab}$ and $\hat{g}_{ab}$. Details of the subleading scalar amplitude may vary as well. Nevertheless, specializing only to conformal Kerr-Schild transformations, meaning that $Y=0$ and $w_a = V k_a$ for some $V$, these amplitudes transform as $\calA_1 \mapsto \Omega^{-1} ( \calA_1 - i \vartheta \calA_0)$. The scalar $\vartheta$ satisfies the transport equation \eqref{varthetaEvolve} and may be interpreted as inducing only a phase shift; cf. \eqref{psiPhase}. While these shifts might be measurable in general, they do not affect, e.g., the averaged stress-energy tensor. 

For electromagnetic fields at one order beyond geometric optics, there is not necessarily any single notion of a corrected wavevector. Such a notion does however exist at least when the leading-order field is linearly polarized, and in that case, the corrected wavevector remains invariant under metric transformations generated by arbitrary $\Omega$, $w_a$, and $Y$. Such transformations can nevertheless act nontrivially on the bare electromagnetic amplitudes. Restricting to conformal Kerr-Schild transformations while allowing for general polarization states, the subleading electromagnetic field is found to experience a kind of generalized phase shift in general; cf. \eqref{fPhase}. At least in the linearly-polarized case, this shift does not affect the averaged stress-energy tensor. For general polarization states, the phase shift vanishes when the shear is trivial or the metric transformation is purely conformal.

One application of our results is that new solutions can be generated from old ones---either in the original spacetime or in new ones. Some examples of this are considered in Sects. \ref{Sect:Examples} and \ref{Sect:Scattering}, where it is shown that even drastic changes in geometry---such as transformations from plane waves to spherical waves---arise from simple operations. One broader result obtained in this way is that there is a sense in which spherical waves are essentially unaffected by (at least static) spherically-symmetric metrics. Similarly, plane-fronted solutions are essentially unaffected by background gravitational waves which propagate in the same direction. Neither of these results are apparent in the coordinate systems most commonly used to describe spherically-symmetric or gravitational-wave geometries.

Indeed, gauge issues can obscure many of the results discussed above. This is true in the sense that an optical field may be preserved in two spacetimes only when the gauge which maps points between those spacetimes is chosen appropriately. However, another kind of gauge fixing is also relevant: The gauge-fixed Maxwell and linearized Einstein equations do not enjoy the same metric-invariance properties as their gauge-agnostic counterparts. In the calculations above, the most complicated aspects of the transformed amplitudes were found to be artifacts of the Lorenz gauge choice. These complications disappear in, e.g., the transformation rules summarized by \eqref{fieldSummary}. More generally, it may be observed that in many contexts, calculations performed in practice often fix a gauge at their outset, and although this affords certain simplifications, it hides others. 

The results of this paper may be extended in various ways. One of the more interesting directions would perhaps be to translate them to other areas of physics. Two straightforward examples could involve i) the propagation of sound through a moving fluid, and ii) the propagation of electromagnetic waves through a nontrivial material. In both of these cases, the spacetime would be physically flat. However, there are mathematical analogies which relate such phenomena (with some restrictions) to Klein-Gordon or vacuum Maxwell fields propagating in effective spacetimes with nontrivial effective metrics \cite{Barcelo2011}. If these analogies are used together with the above invariance results, one might expect to find that acoustic or electromagnetic waves are preserved by certain transformations. However, these transformations would be interpreted as acting not on the physical metric, but on the properties of the underlying material. High-frequency fields would thus be seen to propagate identically in structures with widely-varying characteristics, a result which could help to simplify the design of, e.g., new meta-materials.

\ack

I thank Brien Nolan and Marius Oancea for helpful discussions.

\appendix

\section{Metric transformations and their properties}
\label{app:Metric}

Metric transformations $g_{ab} \mapsto \hat{g}_{ab}$ in which $\hat{g}_{ab}$ is given by \eqref{gTilde} play a central role in this paper. We now derive geometrical properties of these transformations and interpret them as compositions of simpler transformations.

\subsection{Extended Kerr-Schild and conformal transformations}

One of the primary building blocks of the general metric transformations considered here can be described as an extended Kerr-Schild transformation. While similar terminology has been used in the literature to refer to somewhat different concepts \cite{Llosa2009, Ett2010}, we use it here it to refer to
\begin{equation}
	g_{ab} \mapsto \mathbb{K}_{ab}(g;v,w) \equiv g_{ab} + v_{(a} w_{b)},
	\label{Kdef}
\end{equation}
where at least one of $v_a$ or $w_a$ is null with respect to $g_{ab}$. Suppose for definiteness that it is $v_a$ with this property, so $g^{ab} v_a v_b = 0$. The Sherman-Morrison formula for inverting matrices with rank-1 perturbations may then be used to invert $\mathbb{K}_{ab}$; letting $g^{-1}(v,w) = g^{ab} v_a w_b$,
\begin{equation}
	g^{ab} \mapsto g^{ab} -  \frac{ g^{a(c} g^{d)b} v_{c} }{  1 + \frac{1}{2} g^{-1} (v, w) } \left[ w_{d} -  \left( \frac{\frac{1}{4} g^{-1} (w, w) }{ 1 + \frac{1}{2} g^{-1} (v, w)  } \right) v_{d} \right] 
	\label{invPert}
\end{equation}
for arbitrary extended Kerr-Schild perturbations. Moreover, the volume element $\epsilon[g]_{abcd}$ associated with $g_{ab}$ may be shown to transform via
\begin{equation}
	\epsilon[g]_{abcd} \mapsto \epsilon[\mathbb{K}(g;v,w)]_{abcd} = [1 + \tfrac{1}{2} g^{-1}(v, w) ] \epsilon[g]_{abcd}
	\label{detPert}
\end{equation}
when $v_a$ is null. This can be applied to determine how the divergence of an arbitrary vector field $Z^a$ transforms under extended Kerr-Schild transformations: Using $\nabla[g]_a$ to denote the covariant derivative associated with $g_{ab}$,
\begin{equation}
	\nabla[\mathbb{K}(g;v,w)]_a Z^a = \nabla[g]_a Z^a + \mathcal{L}_Z \ln [1 + \tfrac{1}{2} g^{-1}( v, w) ].
	\label{divXform}
\end{equation}
A similar result also holds for the divergence of an arbitrary bivector $Z^{ab} = Z^{[ab]}$.	

An ordinary Kerr-Schild transformation is one in which $v_a$ and $w_a$ are proportional and null. Suppose in particular that $v_a = V w_a$, in which case \eqref{invPert}, \eqref{detPert}, and \eqref{divXform} reduce to
\begin{equation}
\begin{split}
	\begin{gathered}
	g^{ab} \mapsto g^{ab} - V g^{ac} g^{bd} w_c w_d
, \qquad \epsilon[g]_{abcd} \mapsto \epsilon[g]_{abcd},
\\
	\nabla[\mathbb{K}(g;w,V w)]_a Z^a = \nabla[g]_a Z^a.
	\end{gathered}
\end{split}
	\label{KSprops}
\end{equation}
Linearly-perturbed inverses are thus exact and volume elements and divergences are preserved.

Besides the extended Kerr-Schild metric transformations $\mathbb{K}_{ab}(g;v,w)$, it is also convenient to consider conformal transformations which act on metrics via
\begin{equation}
	g_{ab} \mapsto \mathbb{C}_{ab}(g;\Omega) \equiv \Omega^2 g_{ab}.
	\label{confDef}
\end{equation}
It is immediately clear that
\begin{equation}
	g^{ab} \mapsto \Omega^{-2} g^{ab}, \qquad \epsilon[g]_{abcd} \mapsto \Omega^4 \epsilon[g]_{abcd},
	\label{confPert}
\end{equation}
and a short calculation confirms that
\begin{equation}
	\nabla[\mathbb{C}(g;\Omega)]_a Z^a = \nabla[g]_a Z^a + 4 \mathcal{L}_Z \ln \Omega.
	\label{divConf}
\end{equation}

\subsection{Compositions of metric transformations}

A large class of interesting metrics may be generated by composing the transformations $\mathbb{C}_{ab}$ and $\mathbb{K}_{ab}$. In particular, we now show that the metrics \eqref{gTilde}, which preserve optical rays associated with a given $k_a$, may be generated by acting on a background $g_{ab}$ with three instances of $\mathbb{K}_{ab}$ and one of $\mathbb{C}_{ab}$. 

To see this, first fix a complex tetrad \eqref{tetrad} which is null and normalized in the standard way with respect to $g_{ab}$. Next, consider the ordinary Kerr-Schild transformation
\begin{equation}
	g_{ab} \mapsto g^{1}_{ab} \equiv \mathbb{K}_{ab} ( g; m, \bar{Y} m )
	\label{gm}
\end{equation}
generated by the complex null basis element $m_a$. Here, $\bar{Y}$ is the complex conjugate of an arbitrary, possibly-complex scalar field. It follows immediately from \eqref{KSprops} that if $k_a$ is null with respect to $g_{ab}$, it remains null under all such transformations. However, physical metrics must be real and $g_{ab}^1$ is not. This may be remedied by applying a second Kerr-Schild transformation, but in a direction which is in some way related to $\bar{m}_a$. Crucially, this second transformation cannot be generated by $\bar{m}_a$ itself, because $g^1_{ab} \mapsto g^1_{ab} + (\ldots) \bar{m}_a \bar{m}_b$ is not a Kerr-Schild transformation; although $\bar{m}_a$ is null with respect to $g_{ab}$, it is not null with respect to $g^1_{ab}$.

A second Kerr-Schild transformation may instead be generated by using $\bar{m}_a + \frac{1}{2} \bar{Y} m_a$, which is null with respect to $g^1_{ab}$. More precisely, it is convenient to let
\begin{equation}
g_{ab}^2 \equiv \mathbb{K}_{ab} \left( g^1; \bar{m} + \tfrac{1}{2} \bar{Y} m, \frac{ Y  }{ 1- \frac{1}{4} |Y|^2 } (\bar{m} + \tfrac{1}{2} \bar{Y} m)\right),
\label{gPerp}
\end{equation}
where it has been assumed that $|Y|^2 \neq 4$. The scalar $Y/ (1-\frac{1}{4} |Y|^2)$ which appears here is chosen to ensure that $g_{ab}^2$ is real. Using \eqref{Kdef} and \eqref{gm}, 
\begin{equation}
	g_{ab}^2 = g_{ab} + \frac{\bar{Y} m_a m_b + |Y|^2 m_{(a} \bar{m}_{b)} + Y \bar{m}_a \bar{m}_b }{1-\frac{1}{4} |Y|^2 }.
	\label{g2Def}
\end{equation}
It follows immediately from \eqref{KSprops} and \eqref{gPerp} that since $k_a$ is null with respect to $g_{ab}^1$ (and of course $g_{ab}$),  it is also null with respect to $g_{ab}^2$. Note that $g^2_{ab}$ is a composition of two Kerr-Schild transformations acting on $g_{ab}$, and that such compositions result in metric perturbations which are distinct from naive sums of Kerr-Schild terms such as, e.g., $\bar{Y} m_a m_b + Y \bar{m}_a \bar{m}_b$; the $m_{(a} \bar{m}_{b)}$ cross term in \eqref{g2Def} is essential to retaining simple transformation laws for inverses and volume elements, and also for preserving the optical rays.

Two more transformations are required to generate all metrics $\hat{g}_{ab}$ with the form \eqref{gTilde}. These are apparent essentially by inspection, and may be written as
\begin{align}
	\hat{g}_{ab} = \mathbb{C}_{ab}( \mathbb{K}(g^2;k,w) ; \Omega ) = \Omega^2 ( g_{ab}^2 + k_{(a} w_{b)}) . 
	\label{gTildeComp}
\end{align}
Recalling that $k_a$ is considered to be given here, the transformations $g_{ab} \mapsto \hat{g}_{ab}$ are generated by the (nearly) arbitrary real covector $w_a$, the real scalar $\Omega$, and the complex scalar $Y$. Together, these constitute $4+1+2 = 7$ real functions. While different basis elements $m_a$ and $\bar{m}_a$ may be chosen, this freedom has no physical significance; it merely shifts the deformation functions $\Omega$, $w_a$, and $Y$ needed to map between a given pair of  metrics.

\subsection{Properties of $\hat{g}_{ab}$}

Eqs. \eqref{gm}, \eqref{gPerp}, and \eqref{gTildeComp} clearly show how the $\hat{g}_{ab}$ given by \eqref{gTilde} can be generated by applying to $g_{ab}$ two Kerr-Schild transformations, an extended Kerr-Schild transformation, and a conformal transformation. This decomposition may be used to easily compute volume elements, inverses, and divergences.

To begin, consider $g_{ab}^1$ and $g_{ab}^2$. Recalling from \eqref{KSprops} that a single Kerr-Schild transformation does not affect volume elements, two such transformations do not do so either. The same comments also apply for the divergence of an arbitrary vector field $Z^a$. Hence,
\begin{equation}
	\epsilon[g^2]_{abcd} = \epsilon[g^1]_{abcd} = \epsilon[g]_{abcd}, \qquad \nabla[g^2]_a Z^a = \nabla[g^1]_a Z^a = \nabla[g]_a Z^a.
	\label{epsPerp}
\end{equation}
Moreover, two applications of \eqref{KSprops} show that the inverse of $g_{ab}^2$ is
\begin{equation}
	g^{ab}_2 = g^{ab} - \frac{\bar{Y} m^a m^b - |Y|^2 m^{(a} \bar{m}^{b)} + Y \bar{m}^a \bar{m}^b }{1-\frac{1}{4} |Y|^2 },
	\label{gPerpInv}
\end{equation}
where indices on $m_a$ and $\bar{m}_a$ have been raised using $g^{ab}$. 

These results may be combined with \eqref{detPert}, \eqref{confPert}, and  \eqref{gTildeComp} to yield the volume element associated with the fully-transformed metric $\hat{g}_{ab}$. Letting $\hat{\epsilon}_{abcd} \equiv \epsilon[\hat{g}]_{abcd}$, $\epsilon_{abcd} \equiv \epsilon[g]_{abcd}$, and $k \cdot w = g^{ab} k_a w_b$,
\begin{equation}
	\hat{\epsilon}_{abcd} = \Omega^4 ( 1 + \tfrac{1}{2} k \cdot w)
	 \epsilon_{abcd}.
	 \label{epsHat}
\end{equation} 
Furthermore, it follows from \eqref{invPert} and \eqref{confPert} that
\begin{equation}
	\hat{g}^{ab} = \Omega^{-2} \left( g^{ab}_2 -g_2^{a(c} g^{d)b}_2  k_{c} W_{d} \right),
	\label{gInvHat}
\end{equation}
where 
\begin{equation}
	W_a \equiv \frac{1}{ 1 + \frac{1}{2} k \cdot w } \left[ w_a -  \left( \frac{  \frac{1}{4}  g^{-1}_2( w, w) }{ 1 + \frac{1}{2} k \cdot w } \right)  k_a \right].
	\label{Wdef}
\end{equation}
If $g_{ab}$ is itself non-singular, and if the various deformation functions are themselves well-behaved, $\hat{g}_{ab}$ remains non-singular and invertible as long as
\begin{equation}
	|Y|^2 \neq 4, \qquad k \cdot w \neq - 2, \qquad \Omega \neq 0.
	\label{badParams}
\end{equation}
Assuming that these conditions are satisfied, the divergence of an arbitrary vector field with respect to $\hat{g}_{ab}$ may be computed by combining \eqref{divXform} and \eqref{divConf} with \eqref{epsPerp}: Letting $\hat{\nabla}_a \equiv \nabla[\hat{g}]_a$ and $\nabla_a \equiv \nabla[g]_a$,
\begin{align}
	\hat{\nabla}_a Z^a = \nabla_a Z^a + \mathcal{L}_Z \ln[ \Omega^4 ( 1 + \tfrac{1}{2} k \cdot w ) ].
	\label{divHat}
\end{align}
The same result could also have been obtained using \eqref{epsHat}. 

\subsection{A transformed null tetrad}

It is convenient for many purposes to consider a tetrad of covectors \eqref{tetrad} for which the only non-vanishing inner products with respect to $g_{ab}$ are given by \eqref{tetradNorm}. With this in hand, it is possible to construct a transformed counterpart
\begin{equation}
	(k_a, \hat{n}_a, \hat{m}_a, \hat{\bar{m}}_a)
	\label{tetradHat}
\end{equation}
which is normalized in the same way, but with respect to $\hat{g}_{ab}$ instead of $g_{ab}$. The main additional constraint is that the first elements of the transformed and background tetrads are required to be identical, which is possible because (by construction) $k_a$ is null with respect to both metrics. 

Although different transformations of $m_a$ and $\bar{m}_a$ are possible, the possibilities considered here are parameterized using a real $\theta$ and a complex $l$. In particular,
\begin{equation}
	\hat{m}_a \equiv \frac{ \Omega e^{i\theta} }{ ( 1 - \frac{1}{4} |Y|^2 )^{ \tfrac{1}{2}} } ( m_a + \tfrac{1}{2} Y \bar{m}_a + l k_a).
	\label{mTilde}
\end{equation}
It may be verified using \eqref{gPerpInv}, \eqref{gInvHat}, and \eqref{Wdef} that the algebraic constraints on the tetrad are satisfied for any $\theta$ and $l$: $\hat{g}^{-1}( \hat{m}, \hat{m}) = \hat{g}^{-1} (k, \hat{m}) = 0$ and $\hat{g}^{-1} ( \hat{m}, \hat{\bar{m}} ) = 1$. Although it is rarely needed, a correctly-normalized replacement for $n_a$ may be shown to have the form
\begin{align}
	\hat{n}_a = \Omega^2 \left\{ (1+\tfrac{1}{2} k \cdot w) n_a + \Re \left[ \left( \frac{ 2 \bar{l} + \bar{Y} l }{ 1- \frac{1}{4} |Y|^2 } - \bar{m} \cdot w \right) m_a \right] + d k_a \right\}
	\label{nTilde}
\end{align}
for some real $d$. Even without specifying $d$, this expression guarantees that $\hat{g}^{-1} (\hat{n}, \hat{m}) = 0$ and $\hat{g}^{-1} (k, \hat{n}) = -1$. We nevertheless assume that $\lambda$ is chosen to additionally ensure that $\hat{g}^{-1}(\hat{n},\hat{n}) =0$, which is always possible.

Hatted tetrads which are constructed using \eqref{mTilde} and \eqref{nTilde} serve as appropriately-normalized transformations of the background tetrad \eqref{tetrad}. However, the scalars $\theta$ and $l$ which appear in their definition are thus-far unconstrained. Constraints on these scalars do appear if it is supposed that the tetrads are not only normalized, but also that they are parallel transported along the optical rays (by their respective metrics). By construction, $k_a$ is parallel transported with respect to both $g_{ab}$ and $\hat{g}_{ab}$. If $m_a$ is also parallel transported with respect to $g_{ab}$, parallel transport of $\hat{m}_a$ with respect to $\hat{g}_{ab}$ may be shown to imply that $\theta$ and $l$ satisfy certain transport equations along the rays. The first of these transport equations is
\begin{equation}
	\mathcal{L}_k \theta  = \frac{ \Im  ( \bar{\sigma} Y ) + \frac{1}{4} |Y|^2 \mathcal{L}_k \arg Y   }{ 1 - \frac{1}{4} |Y|^2 } ,
	\label{theta}
\end{equation}
where 
\begin{equation}
	\sigma \equiv - m^a m^b \nabla_a k_b
	\label{sigDef}
\end{equation} 
denotes the complex shear of the optical rays with respect to $g_{ab}$. Tetrad rotations associated with nontrivial variations in $\theta$ thus depend only on the deformation function $Y$; they are unaffected by those portions of the metric transformation which depend on $\Omega$ or $w_a$.

The transport equation for $l$ which guarantees the parallel transport of $\hat{m}_a$ is complicated. Although we omit it in its most general form, this equation simplifies considerably for conformal Kerr-Schild transformations in which $Y=0$ and $w_a = V k_a$; in those cases,
\begin{equation}
	\mathcal{L}_k ( \Omega e^{i \theta} l ) = - \mathcal{L}_m \Omega
	\label{lEvolve} 
\end{equation}
and $\mathcal{L}_k \theta = 0$. While this is independent of the Kerr-Schild portion of the transformation generated by $V$, it does depend on $\Omega$. Nontrivial conformal transformations are thus accompanied by nontrivial forms for $l$, at least if $\mathcal{L}_m \Omega \neq 0$. Nevertheless, it is rarely necessary to compute $l$.

\subsection{Shear}

The shear $\sigma$ which is defined by \eqref{sigDef} plays an important role in finite-wavelength corrections to geometric optics. It is therefore of interest to understand how it transforms when the ray congruence is fixed while the metric is deformed. If $m_a$ and $\hat{m}_a$ are parallel transported with respect to the appropriate metrics, the above expressions may be used to show that for arbitrary $\Omega$, $w_a$, and $Y$, 
\begin{equation}
	\hat{\sigma} = \frac{ \Omega^{-2} e^{2 i \theta } }{  (1-\frac{1}{4} |Y|^2)(1+\frac{1}{2} k \cdot w) }  \left( \sigma - \tfrac{1}{4} Y^2 \bar{\sigma} - \tfrac{1}{2} \mathcal{L}_k Y \right) .
	\label{sigxForm}
\end{equation}
One implication is that if the congruence of interest is shear-free with respect to $g_{ab}$, it remains shear-free for all $\hat{g}_{ab}$ in which $\mathcal{L}_k Y = 0$; the deformation scalars $\Omega$ and $w_a$ cannot be used to generate shear if there is initially none. If transformations are considered only from vacuum metrics to vacuum metrics in which the shear initially vanishes, this and the Goldberg-Sachs theorem \cite{ExactSolns} imply that both the original and transformed metrics must be algebraically special. It may also be shown that the (background) shear satisfies the evolution equation $\mathcal{L}_k \sigma = R_{abcd} k^a m^b k^c m^d - \sigma (\nabla \cdot k)$.

\section{Notation and conventions}
\label{app:Notation}

It is assumed that all metrics are four dimensional and have signature $(-+++)$. Lower-case Latin letters $a,b,\ldots$ are used for ordinary abstract indices, upper-case Latin letters $A,B, \ldots$ for abstract multi-indices, and Greek letters $\mu,\nu, \ldots$ for coordinate indices. The Riemann curvature tensor is defined such that $2 \nabla_{[a} \nabla_{b]} v_c = \nabla_a \nabla_b v_c - \nabla_b \nabla_a v_c = R_{abc}{}^{d} v_d$ for any $v_a$. Units are used in which $G=c=1$.

We employ different metrics on the same spacetime manifold. When indices are raised or lowered without explicitly specifying the metric involved, it is assumed to be the one denoted by $g_{ab}$. For example, $k_a$ and $w_a$ appear naturally with their indices down, but $k \cdot w \equiv g^{ab} k_a w_b$. Except in some cases where these quantities coincide, operators and other objects associated with $\hat{g}_{ab}$ or $\tilde{g}_{ab}$ are typically adorned with a hat or tilde to distinguish them from counterparts associated with $g_{ab}$. A partial list of the symbols used in the paper is given in Tables \ref{Table2} and \ref{Table1}.

\begin{table}[H]
	\centering
	\begin{tabular}{c | c | c }
	\hline 
	Quantity & Representative equations & Description \\ 
	\hline \hline 
	$\alpha$	& \eqref{A0flat}	&	Waveform\\
	\hline
	$\theta$	&	\eqref{mTilde}, \eqref{theta}	& Angle used to define $\hat{m}_a$ \\
	\hline
	$\Theta$	& \eqref{dsFlat}	&	Coordinate\\
	\hline
	$\vartheta_{A}{}^{B}$	& \eqref{A1Hat}, \eqref{A1HatEM}	& Phase corrections for $\hat{\calA}^1_B$\\
	\hline
	$\mu$	&	\eqref{Mdef}	&	Scalar field mass\\
	\hline 
	$\xi$	&	\eqref{Mdef}	&	Scalar curvature coupling\\
	\hline
	$\xi^a$	&	\eqref{phigTilde}, \eqref{xiEx} & Gauge vector\\
	\hline	 
	$\phi$	& \eqref{diffeo1}, \eqref{diffeo2}, \eqref{phigTilde} 	&	Diffeomorphism\\
	\hline
	$\Phi$	&	\eqref{dsFlat}	&	Coordinate\\
	\hline
	$\Phi_i$	& \eqref{Phi2}, \eqref{Phi0}, \eqref{Phi1} & Electromagnetic Newman-Penrose scalars \\
	\hline
	$\varphi$ & \eqref{WKB}, \eqref{eikonal} & Eikonal \\ 
	\hline
	$\varphi^\mathrm{cor}$	& \eqref{Kcorrect}	& Corrected eikonal\\
\hline 
$\chi_B$ &	\eqref{eExpand}, \eqref{eExpandGrav} & Polarization component \\
\hline
$\psi_B$ &	\eqref{WKB}	&	(Complexified) high-frequency field\\
\hline
$\Psi_B$	& \eqref{RePsi} & Real high-frequency field\\
\hline
$\delta \Psi_i$	& \eqref{Psi4}, \eqref{Psi0}, \eqref{Psi2}, \eqref{Psi3}	&	Gravitational Newman-Penrose scalars\\
\hline
$\omega$	&	\eqref{WKB}	& Frequency parameter\\
\hline
$\Omega$ & \eqref{gTilde} & Conformal factor \\ 
\hline 
	\end{tabular} 
	\caption{Notation (Greek)}
	\label{Table2}
\end{table}

\begin{table}[H]
	\centering
	\begin{tabular}{c | c | c }
	\hline 
	Quantity & Representative equations & Description \\ 
	\hline \hline 
	$\calA^n_B$	& \eqref{WKB}, \eqref{xPort}, \eqref{gaugeEM}, \eqref{gaugeGrav} &	$n$th-order amplitude\\
	\hline
	$\mathbb{C}_{ab}$ & \eqref{confDef}	&	Conformal transformation\\
	\hline
	$e_\pm$ & \eqref{eExpand}, \eqref{eExpandGrav}	&	Polarization components\\
	\hline
	$e_B$ &\eqref{A0s}, \eqref{eExpand}, \eqref{eExpandGrav} &	Polarization tensor \\
	\hline
	$f_{ab}$ & \eqref{fExpand}	& (Complexified) electromagnetic field\\
	\hline
	$F_{ab}$	&	$\Re f_{ab}$ & Real electromagnetic field\\
	\hline
	$\mathcal{F}^n_{ab}$ & \eqref{fExpand}, \eqref{Fcal0} & $n$th-order electromagnetic field coefficient\\
	\hline
$g_{ab}$ & & Background metric  \\
	\hline
	$\hat{g}_{ab}$	& \eqref{gTilde}, \eqref{gTildeSimp}, \eqref{gConfKS}, \eqref{gTildeComp} &	Transformed metric\\ 
	\hline
	$\tilde{g}_{ab}$	&	\eqref{gPert}  & Perturbed metric	\\
	\hline
	$J^a_0$	& \eqref{J0sc}	&	Conserved flux\\
	\hline
	$k_a$ & \eqref{eikonal} & Wavevector \\ 
	\hline
	$k_a^{\mathrm{cor}}$ & \eqref{Kcorrect}, \eqref{Kpm}, \eqref{KpmGrav}	& Corrected wavevector \\
	\hline
	$\mathbb{K}_{ab}$	& \eqref{Kdef}	&	Extended Kerr-Schild transformation\\
	\hline
	$l$	&	\eqref{mTilde}, \eqref{lEvolve}	& Correction in $\hat{m}_a$\\
\hline 
$L$	&	\eqref{LDef}	&	Transport operator\\
\hline
$\mathcal{L}$	&	& Lie derivative\\
\hline
$m_a$ & \eqref{tetrad}, \eqref{tetradNorm}, \eqref{mTilde} & Complex basis element \\ 
\hline
$M$ & \eqref{hNewt}	& Mass\\
\hline
$n_a$	&	\eqref{tetrad}, \eqref{tetradNorm}, \eqref{nTilde}	&	Real basis element\\
\hline 
$O_n$	&	\eqref{ODef}	& $n$th-order optical fields\\
\hline
$r$	& $r=\sqrt{x^2+y^2+z^2} = \frac{1}{2}(v-u)$	&	Coordinate\\
\hline
$\delta r_{abcd}$ &	\eqref{RiemExpand} & (Complexified) curvature perturbation\\
\hline
$\delta R_{abcd}$ & $\Re \delta r_{abcd}$ & Real curvature perturbation\\
\hline
$\mathcal{R}^n_{abcd}$	& \eqref{RiemExpand} & $n$th-order curvature coefficient\\
\hline
	$s$	&	&	Spin of the field (0, 1, 2)\\
	\hline
	$t$	&	$t = \frac{1}{2} (v+u)$ &	Coordinate\\
	\hline
	$\langle T_{ab} \rangle$	& \eqref{Texpand}	&	Averaged stress-energy tensor\\
	\hline
	$\langle \delta T_{abcd} \rangle$	& \eqref{BelRobinson}	&	Averaged Bel-Robinson perturbation\\
	\hline
	$u$	&	\eqref{dsFlat}, \eqref{eikonalFlat}	&	Coordinate\\
	\hline
	$v$	&	\eqref{dsFlat}	&	Coordinate\\
	\hline	
	$V$	&	\eqref{gConfKS}	&	Kerr-Schild deformation\\
	\hline	
	$w_a$ & \eqref{gTilde}, \eqref{gTildeComp}  & Metric deformation \\ 
	\hline 
	$Y$ & \eqref{gTilde}, \eqref{g2Def} & Transverse metric deformation \\ 
	\hline 
	$z$	&	\eqref{zDef}	& Splitting of stress-energy eigenvectors\\
		&					& (or a coordinate in Sect. \ref{Sect:Scattering}) \\
	\hline
	\end{tabular} 
	\caption{Notation (Latin)}
	\label{Table1}
\end{table}

\section*{References}

\bibliographystyle{iopart-num}
\bibliography{refsLens}

\providecommand{\newblock}{}
\begin{thebibliography}{10}
\expandafter\ifx\csname url\endcsname\relax
  \def\url#1{{\tt #1}}\fi
\expandafter\ifx\csname urlprefix\endcsname\relax\def\urlprefix{URL }\fi
\providecommand{\eprint}[2][]{\url{#2}}

\bibitem{SchneiderEhlers}
Schneider P, Ehlers J and Falco E~E 1992 {\em Gravitational lenses\/}
  (Springer)

\bibitem{BartelmannRev}
Bartelmann M 2010 {\em \CQG\/} {\bf 27} 233001

\bibitem{darkMatter}
Massey R, Kitching T and Richard J 2010 {\em Rep. Prog. Phys.\/} {\bf 73}
  086901

\bibitem{InverseProbs}
{Moura Neto} F~D and da~Silva~Neto A~J 2012 {\em An introduction to inverse
  problems with applications\/} (Springer)

\bibitem{BatemanConf}
Bateman H 1910 {\em Proc. Lond. Math. Soc.\/} {\bf 8} S2--223

\bibitem{Wald}
Wald R~M 1984 {\em General relativity\/} (University of Chicago Press)

\bibitem{Kastrup2008}
Kastrup H 2008 {\em Ann. Phys. (Leipzig)\/} {\bf 17} 631

\bibitem{HarteEMKS}
Harte A~I 2017 {\em Phys. Rev. Lett.\/} {\bf 118} 141101

\bibitem{HarteOptics1}
Harte A~I 2019 {\em Gen. Rel. Grav.\/} {\bf 51} 14

\bibitem{BornWolf}
Born M and Wolf E 1999 {\em Principles of optics\/} (Cambridge University
  Press)

\bibitem{Keller}
Keller J~B and Lewis R~M 1995 Asymptotic methods for partial differential
  equations: The reduced wave equation and {M}axwell's equations {\em Surveys
  in applied mathematics\/} ed Keller J~B, McLaughlin D~W and Papanicolaou G~C
  (Springer) p~1

\bibitem{EhlersGeoOptics}
Ehlers J 1967 {\em Z. Naturforsch.\/} A {\bf 22} 1328

\bibitem{Anile}
Anile A~M 1976 {\em J. Math. Phys.\/} {\bf 17} 576

\bibitem{Isaacson1}
Isaacson R~A 1968 {\em Phys. Rev.\/} {\bf 166} 1263

\bibitem{DolanGeoOptics1}
Dolan S~R 2018 {\em Int. J. Mod. Phys.\/} D {\bf 27} 1843010

\bibitem{ThorneBlandford}
Thorne K~S and Blandford R~D 2017 {\em Modern classical physics: Optics,
  fluids, plasmas, elasticity, relativity, and statistical physics\/}
  (Princeton University Press)

\bibitem{Guerses}
G\"{u}rses M and G\"{u}rsey F 1975 {\em J. Math. Phys.\/} {\bf 16} 2385

\bibitem{Xanthopoulos1978}
Xanthopoulos B~C 1978 {\em J. Math. Phys.\/} {\bf 19} 1607

\bibitem{ExactSolns}
Stephani H, Kramer D, MacCallum M, Hoenselaers C~L~U and Herlt E 2009 {\em
  Exact solutions of Einstein's field equations\/} (Cambridge University Press)

\bibitem{PenroseLimit}
Penrose R 1976 Any space-time has a plane wave as a limit {\em Differential
  geometry and relativity\/} (Springer) p 271

\bibitem{Blau}
Blau M, Frank D and Weiss S 2006 {\em Class. Quantum Grav.\/} {\bf 23} 3993

\bibitem{Llosa2009}
Llosa J and Carot J 2009 {\em Class. Quantum Grav.\/} {\bf 26} 055013

\bibitem{HarteXKS}
Harte A~I 2014 {\em Phys. Rev. Lett.\/} {\bf 113} 261103

\bibitem{OptGeo}
Robinson I and Trautman A 1989 Optical geometry {\em Proceedings of the XI
  Warsaw symposium on elementary particle physics\/} (World Scientific) p 454

\bibitem{MarsEhlersGroup}
Mars M 2001 {\em Class. Quantum Grav.\/} {\bf 18} 719

\bibitem{PoissonRev}
Poisson E, Pound A and Vega I 2011 {\em Living Rev. Relativ.\/} {\bf 14}

\bibitem{Ellis1985}
Ellis G, Nel S, Maartens R, Stoeger W and Whitman A 1985 {\em Phys. Rep.\/}
  {\bf 124} 315

\bibitem{Gasperini2011}
Gasperini M, Marozzi G, Nugier F and Veneziano G 2011 {\em J. Cosm. Astropart.
  Phys.\/} {\bf 07} 008

\bibitem{Fleury2016}
Fleury P, Nugier F and Fanizza G 2016 {\em J. Cosm. Astropart. Phys.\/} {\bf
  06} 008

\bibitem{spinHallRev}
Oancea M~A, Paganini C~F, Joudioux J and Andersson L {\em arXiv:1904.09963\/}

\bibitem{SenovillaKS}
Coll B, Hildebrandt S~R and Senovilla J~M~M 2001 {\em Gen. Rel. Grav.\/} {\bf
  33} 649

\bibitem{SenovillaBiconformal}
Garc{\'{\i}}a-Parrado A and Senovilla J~M~M 2004 {\em Class. Quantum Grav.\/}
  {\bf 21} 2153

\bibitem{confFRW}
Ibison M 2007 {\em J. Math. Phys.\/} {\bf 48} 122501

\bibitem{confFRW2}
Iihoshi M, Ketov S~V and Morishita A 2007 {\em Prog. Th. Phys.\/} {\bf 118} 475

\bibitem{confKS}
Mitskievich N~V and Horsk{\'{y}} J 1996 {\em Class. Quantum Grav.\/} {\bf 13}
  2603

\bibitem{GriffithsExact}
Griffiths J~B and Podolsk\'{y} J 2010 {\em Exact space-times in Einstein's
  general relativity\/} (Cambridge University Press)

\bibitem{HarteVines}
Harte A~I and Vines J 2016 {\em Phys. Rev.\/} D {\bf 94} 084009

\bibitem{Barcelo2011}
Barcel{\'{o}} C, Liberati S and Visser M 2011 {\em Living Rev. Relativ.\/} {\bf
  14} 3

\bibitem{Ett2010}
Ett B and Kastor D 2010 {\em Class. Quantum Grav.\/} {\bf 27} 185024

\end{thebibliography}

\end{document}